\documentclass[aps,prl,reprint,superscriptaddress,floatfix]{revtex4-2}
\usepackage{amsmath,amssymb,graphicx,color,verbatim,ulem,braket,tabularx,multirow,array,diagbox,colortbl,hhline,float}
\usepackage[colorlinks,linkcolor=blue,citecolor=blue,urlcolor=blue]{hyperref}
\usepackage{physics}
\usepackage{bm}
\usepackage{blindtext}
\newcommand\startsupplement{%
  \setcounter{figure}{0} 
  \renewcommand{\thefigure}{S\arabic{figure}} 
  \setcounter{table}{0}
  \renewcommand{\thetable}{S\arabic{table}}
  \setcounter{equation}{0}
  \renewcommand{\theequation}{S\arabic{equation}}
  \setcounter{section}{0}
  \renewcommand{\thesection}{S\arabic{section}}
}

\graphicspath{{Figures/}}

\begin{document}

\title{Thermal Signatures of the Slater-Mott Crossover in the Hubbard Model: \\
From Double Occupancy to Antiferromagnetic Correlation Length}

\author{Mingzhong Lu}
\affiliation{Department of Modern Physics, University of Science and Technology of China, Hefei, Anhui 230026, China}
\affiliation{Institute of Modern Physics, Northwest University, Xi'an 710127, China}

\author{Yu-Feng Song}
\affiliation{Hefei National Laboratory for Physical Sciences at Microscale and Department of Modern Physics, University of Science and Technology of China, Hefei, Anhui 230026, China}
\affiliation{Institute of Modern Physics, Northwest University, Xi'an 710127, China}

\author{Youjin Deng}
\email{yjdeng@ustc.edu.cn}
\affiliation{Department of Modern Physics, University of Science and Technology of China, Hefei, Anhui 230026, China}
\affiliation{Hefei National Laboratory for Physical Sciences at Microscale and Department of Modern Physics, University of Science and Technology of China, Hefei, Anhui 230026, China}
\affiliation{Hefei National Laboratory, Hefei 230088, China}

\author{Yuan-Yao He}
\email{heyuanyao@nwu.edu.cn}
\affiliation{Institute of Modern Physics, Northwest University, Xi'an 710127, China}
\affiliation{Shaanxi Key Laboratory for Theoretical Physics Frontiers, Xi'an 710127, China}
\affiliation{Fundamental Discipline Research Center for  Quantum Science and Technology of Shaanxi Province, Xi'an 710127, China}
\affiliation{Hefei National Laboratory, Hefei 230088, China}

\begin{abstract}
The interaction-driven crossover from a Slater insulator to a Mott insulator in the N\'{e}el-ordered ground state of the Hubbard model is a fundamental paradigm of strongly correlated electrons, yet its quantitative characterization has remained elusive. Here we establish a clear and experimentally accessible thermal criterion for this crossover via the sign change of the temperature derivative of double occupancy, $(\partial D/\partial T)_U$, near zero temperature. In the Slater regime, $(\partial D/\partial T)_U>0$ reflects the major role of charge fluctuations; in the Mott regime, the anomalous $(\partial D/\partial T)_U<0$, a manifestation of the Pomeranchuk effect, signals the dominance of low-energy spin superexchange physics. Using exact diagonalization and {\it numerically exact} quantum Monte Carlo simulations, we demonstrate that this criterion determines the crossover boundary at $U_{\rm cross}/t=4.0(2)$ for the half-filled two-dimensional Hubbard model. Furthermore, we yield a consistent boundary independently from the maximum in the antiferromagnetic correlation length, which also arises from the superexchange physics. These two thermal signatures are theoretically unified through the local minimum of thermal entropy versus interaction $U$ at low temperatures. Our results offer a direct, measurable, and physically intuitive framework to identify the Slater-Mott crossover in optical lattice experiments. 
\end{abstract}

\date{\today}
\maketitle

The interplay between the Fermi surface geometry and electron-electron interactions is central to strongly correlated electron systems~\cite{Georges1996,Imada1998,Lee2006}. In the Hubbard model~\cite{Hubbard1963,Kanamori1963,Gutzwiller1963} with nested Fermi surfaces, an infinitesimal repulsion opens a single-particle gap and induces N\'{e}el antiferromagnetic (AFM) long-range order in two and three dimensions (2D and 3D)~\cite{Arovas2022,Qin2022,Hirsch1985}, rendering the ground state an AFM insulator for any $U>0$. However, the underlying mechanism of this insulating state changes fundamentally as $U$ increases. At weak coupling, the insulating behavior arises from Fermi surface nesting and weak N\'{e}el AFM order, with substantial charge fluctuations, defining the Slater insulator~\cite{Slater1951}. At strong coupling, charge fluctuations are suppressed and the low-energy physics is controlled by superexchange between local moments~\cite{Delannoy2005}, which characterize the Mott insulator~\cite{Mott1949,Mott1968}. These two regimes are connected by a smooth crossover~\cite{Imada1998,Lee2006,Junyi2020,Pruschke2003}, whose quantitative and precise characterization remains a considerable challenge.

Existing studies of the Slater-Mott crossover are rather limited and fall into two broad categories. The first approach relies on the distinct spectral features in the Slater and Mott regimes~\cite{Pruschke2003,Borejsza2003,Fratino2017,Wang2019}. They include the fine and delicate structures in the single-particle spectral function obtained from approximate theories or numerical methods~\cite{Pruschke2003,Borejsza2003,Fratino2017}. These features, however, often lack a clear criterion for locating the crossover. In addition, linearly extrapolating the spectral gap (i.e., $\Delta_{sp}\propto U$)~\cite{Wang2019} captures only the physics deep in the Mott regime. A second approach examines the energetics of the Hubbard model to infer the crossover~\cite{Fratino2017,Gull2008,Rohringer2016,vanLoon2018,Kim2020}. Most of these studies focus on contrasting temperature variations of kinetic and potential energies at weak and strong interactions, within dynamical mean-field theory~\cite{Fratino2017,Gull2008,Rohringer2016,vanLoon2018} in the context of the false Mott metal-insulator transition~\cite{Geles2015}, where the quantitative reliability remains limited. Similarly, a recent diagrammatic Monte Carlo study~\cite{Kim2020} associates the Slater insulator with a negative temperature derivative of the kinetic energy. Nevertheless, it is restricted to weak and intermediate interactions, and the connection to the ground state is not clarified. While insightful, these prior studies are largely qualitative and have limited applicability to both various theoretical methods~\cite{Arovas2022,Qin2022} and optical lattice experiments~\cite{Mazurenko2017,Xu2025,Kendrick2025,Bourgund2025,Shao2024,YuXuan2025,Thomas2026}. A simple, quantitatively controlled, and experimentally accessible criterion for the Slater-Mott crossover is therefore still missing. 

In this Letter, we establish such a criterion based on the sign of the temperature derivative of double occupancy, $(\partial D/\partial T)_U$. Near $T=0$, it is positive in the Slater regime and negative in the Mott regime. We validate this criterion using two complementary numerical methods: exact diagonalization (ED) on small systems to reveal the microscopic origin of the sign change, and large-scale auxiliary-field quantum Monte Carlo (AFQMC) simulations to obtain highly precise results of $D$ on extended systems. From the latter, we determine the Slater-Mott crossover boundary at $U_{\rm cross}/t=4.0(2)$ for the 2D Hubbard model. Via the Maxwell relation, the sign change in $(\partial D/\partial T)_U$ is linked to the local minimum in thermal entropy $\boldsymbol{s}(U)$, which marks the strongest AFM spin correlations. This link provides an independent determination of the crossover through the maximum of the AFM correlation length $\xi(U)$ at low temperatures, yielding a consistent estimate of the crossover boundary.

\begin{figure}[t]
\centering
\includegraphics[width=0.981\columnwidth]{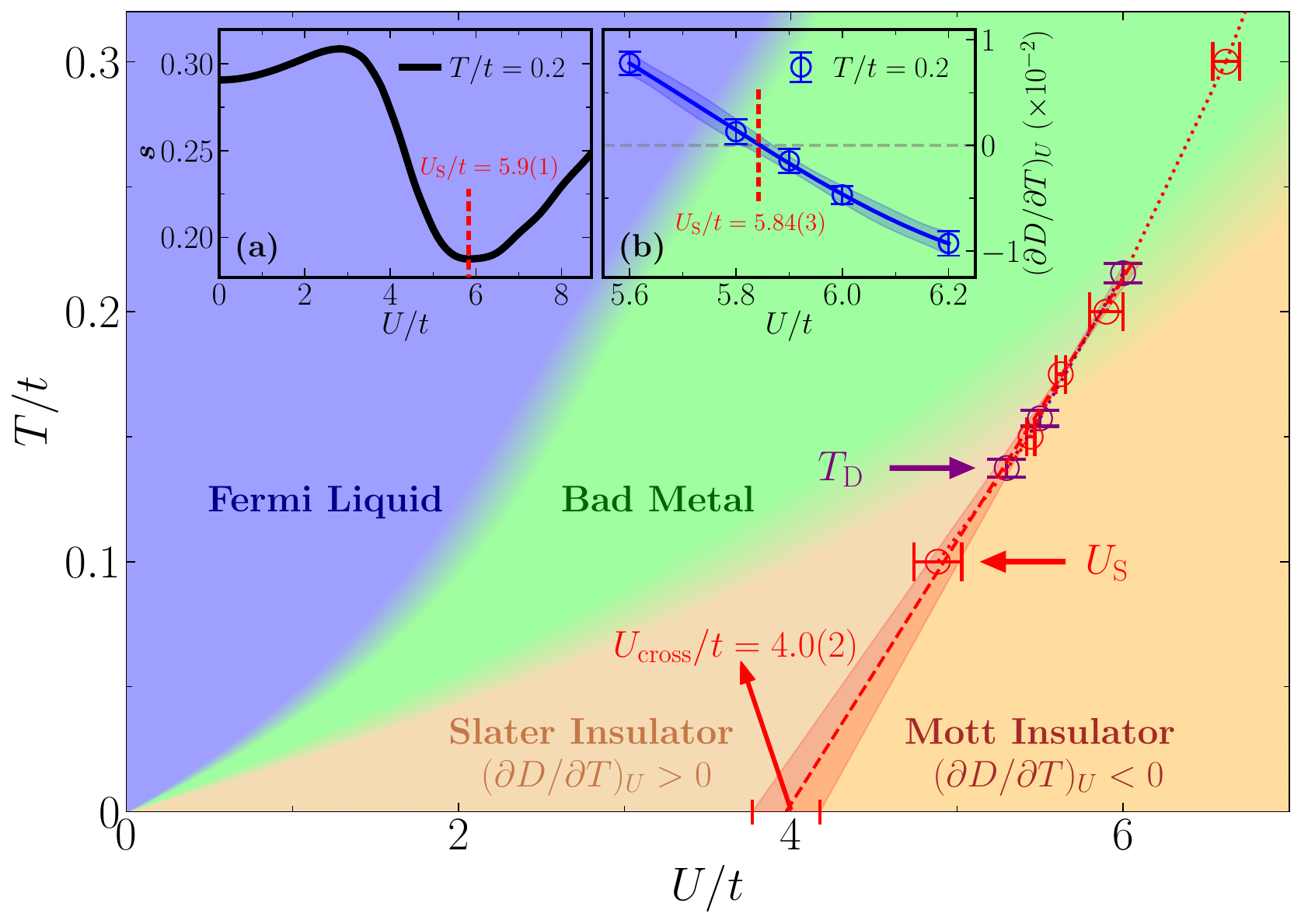}
\caption{
Extrapolation of the thermal signatures at low temperatures to $T=0$ yields $U_{\rm cross}/t=4.0(2)$ as the Slater-Mott crossover boundary for the half-filled square-lattice Hubbard model. The signatures include the local minimum $U_{\rm S}$ in thermal entropy density $\boldsymbol{s}(U)$ at fixed $T/t$ and the minimum $T_{\rm D}$ of double occupancy $D(T)$ at fixed $U/t$ (see Fig.~\ref{fig:Fig03QMCdata}), which are connected via the Maxwell relation in Eq.~(\ref{eq:Maxwell}). A linear fit (red dashed line) to the data is used with the red shaded region indicating the fitting uncertainty. The characteristic feature for the Slater (Mott) regime is $(\partial D/\partial T)_U>0$ [$(\partial D/\partial T)_U<0$] near $T=0$. The background reproduces the finite-$T$ metal-insulator crossover diagram~\cite{Lu2026}, separating the Fermi liquid (blue), bad metal (green), and quasi-AFM insulator (yellow). Inset (a) plots $\boldsymbol{s}(U)$ at $T/t=0.2$, which exhibits a local minimum at $U_{\rm S}/t=5.9(1)$. Inset (b) alternatively determines $U_{\rm S}$ via $(\partial D/\partial T)_U=0$ by applying Eq.~(\ref{eq:Maxwell}), giving $U_{\rm S}/t=5.84(3)$.
}
\label{fig:Fig01Cross}
\end{figure}

We study the half-filled 2D Hubbard model on square lattice described by
\begin{equation}\begin{aligned}
\label{eq:Hubbard}
\hat{H}=\sum_{\mathbf{k}\sigma}\varepsilon_{\mathbf{k}}\hat{c}_{\mathbf{k}\sigma}^{\dagger}\hat{c}_{\mathbf{k}\sigma}^{}
+U\sum_{\mathbf{i}}\Big(\hat{n}_{\mathbf{i}\uparrow}\hat{n}_{\mathbf{i}\downarrow}
-\frac{\hat{n}_{\mathbf{i}\uparrow}+\hat{n}_{\mathbf{i}\downarrow}}{2}\Big),
\end{aligned}\end{equation}
where $\mathbf{k}=(k_x,k_y)$ is the momentum (lattice constant $a=1$), and $\hat{n}_{\mathbf{i}\sigma}=\hat{c}_{\mathbf{i}\sigma}^{\dagger}\hat{c}_{\mathbf{i}\sigma}^{}$ (with $\hat{n}_{\mathbf{i}}=\hat{n}_{\mathbf{i}\uparrow}+\hat{n}_{\mathbf{i}\downarrow}$) is the density operator on the site $\mathbf{i}$ with $\sigma$ ($=\uparrow$ or $\downarrow$) denoting spin. The kinetic energy dispersion is $\varepsilon_{\mathbf{k}}=-2t(\cos k_x+\cos k_y)$ with $t$ as nearest-neighbor hopping strength (set as the energy unit), and $U$ is the on-site repulsion. This model has nested Fermi surfaces and an AFM insulating ground state for any $U>0$, but remains paramagnetic at finite $T$ due to the Mermin-Wagner theorem~\cite{Mermin1966}. We solve this model by employing ED (Lanczos algorithm)~\cite{Dagotto1994,Lin1993,Prelovsek2013,Wu2016} and both ground-state and finite-temperature AFQMC methods~\cite{Blankenbecler1981,Hirsch1983,White1989,Scalettar1991,Congjun2005,Assaad2008,He2019B,Sun2024,Song2025L,*Song2025B,Yuanyao2025,Sugiyama1986,Sorella1989,Duhao2025}, bridging zero- and finite-temperature properties. In AFQMC simulations, we use supercells with $N_s = L^2$ lattice sites ($L$ the linear system size), and apply periodic and twist-averaged boundary conditions (PBC and TABC), where the latter can reduce finite-size effects in computed observables~\cite{Lin2001,Xu2024,Qin2016,Vitali2016}. When applying TABC, we practically take $\sim$$30$ sets of twisted angles to achieve well-converged statistics. Additional algorithmic details of AFQMC methods are provided in the Supplemental Material (SM)~\cite{Suppl}. 

From a physical perspective, although the total particle number $\hat{N}=\sum_{\mathbf{i}}\hat{n}_{\mathbf{i}}$ is conserved in the model~(\ref{eq:Hubbard}), the local density $\hat{n}_{\mathbf{i}}$ is not a conserved quantity, since $[\hat{H},\hat{n}_{\mathbf{i}}]\ne0$. Hence, the local charge density can exhibit quantum fluctuations even in the ground state, despite the presence of a gapped single-particle spectrum and vanishing charge compressibility~\cite{Lu2026}. These fluctuations constitute an intrinsic part of the charge fluctuations in the system. Such local charge fluctuations can be quantitatively characterized by the local charge variance, which is related to the double occupancy $D=N_s^{-1}\sum_{\mathbf{i}}\langle\hat{n}_{\mathbf{i}\uparrow}\hat{n}_{\mathbf{i}\downarrow}\rangle$ via $N_s^{-1}\sum_{\mathbf{i}}\langle(\hat{n}_i-n)^2\rangle=2D$, with $n=N_s^{-1}\langle\hat{N}\rangle=1$ at half filling. Thus, $D$ serves as a direct and quantitative metric of local charge fluctuations in the model~(\ref{eq:Hubbard}). As $U/t$ increases, the monotonically decreasing $D$ in the half-filled Hubbard model~\cite{Lu2026,Song2025L,*Song2025B} directly reflects the progressively suppressed charge fluctuations.

At finite temperatures, as $U/t$ increases, the model~(\ref{eq:Hubbard}) undergoes a smooth metal-insulator crossover (MIC)~\cite{LeBlanc2020,Kim2020,Lu2026} from a Fermi liquid first to a bad metal and then to a quasi-AFM insulator, reproduced from Ref.~\cite{Lu2026} and shown as the background of Fig.~\ref{fig:Fig01Cross}. An important feature of this MIC is the nonmonotonic $U$-dependence of thermal entropy density $\boldsymbol{s}=S/N_s$ (in units of $k_B$)~\cite{Song2025L,*Song2025B,Lu2026}. In Fig.~\ref{fig:Fig01Cross} inset (a), we plot representative results of $\boldsymbol{s}(U)$ at $T/t=0.2$. While the local maximum in $\boldsymbol{s}(U)$ identifies the crossover from the Fermi liquid to bad metal~\cite{Lu2026}, its subsequent local minimum at $U=U_{\rm S}$ essentially signals the dominance of the spin superexchange physics at low temperatures~\cite{Song2025L,*Song2025B,Lu2026}. Moreover, the Maxwell relation 
\begin{equation}\begin{aligned}
\label{eq:Maxwell}
\Big(\frac{\partial\boldsymbol{s}}{\partial U}\Big)_T=-\Big(\frac{\partial D}{\partial T}\Big)_U,
\end{aligned}\end{equation}
reveals that the $U_{\rm S}$ curve corresponds to the temperature line on which the double occupancy $D(T)$ (at fixed $U/t$) exhibits a local minimum, denoted as $T_{\rm D}$. Based on this relation, Fig.~\ref{fig:Fig01Cross} inset (b) provides an alternative determination of $U_{\rm S}$ via $(\partial\boldsymbol{s}/\partial U)_{T}=0$, which translates into $(\partial D/\partial T)_U=0$ and yields a $U_{\rm S}$ value consistent with that in inset (a). We then extrapolate the low-temperature signatures of $T_{\rm D}$ (see Fig.~\ref{fig:Fig03QMCdata}) and $U_{\rm S}$ to $T=0$ and obtain the Slater-Mott crossover boundary at $U_{\rm cross}/t=4.0(2)$. Hence, the $\boldsymbol{s}(U)$ results combined with Eq.~(\ref{eq:Maxwell}) establish the crossover criterion near $T=0$: $(\partial D/\partial T)_U>0$ for the Slater regime and $(\partial D/\partial T)_U<0$ for the Mott regime. The $T_{\rm D}$ and $U_{\rm S}$ data (summarized in the SM~\cite{Suppl}) and their linear fit (red dashed line), along with the crossover criterion, are presented in Fig.~\ref{fig:Fig01Cross}. Other fitting functions tested produce crossover boundaries that clearly conflict with the $D(T)$ results from AFQMC (see Fig.~\ref{fig:Fig03QMCdata}).

Associating the sign change of $(\partial D/\partial T)_U$ with the Slater-Mott crossover is rooted in its connection to Mott-Heisenberg physics, motivated by two insights. First, in the $T<T_{\rm D}$ (or $U>U_{\rm S}$) region of Fig.~\ref{fig:Fig01Cross}, $(\partial D/\partial T)_U<0$, known as the Pomeranchuk effect~\cite{Gorelik2012,GangLi2014,Wietek2021,Qiaoyi2023,Song2025L,*Song2025B,Lu2026}, originates from the virtual hopping~\cite{Song2025L,*Song2025B,Lu2026}, which is the microscopic mechanism underlying spin superexchange~\cite{Delannoy2005}. Second, via Eq.~(\ref{eq:Maxwell}), the sign change is linked to the local minimum $U_{\rm S}$ in $\boldsymbol{s}(U)$. Near $U_{\rm S}$ in the quasi-AFM insulator regime, the charge entropy is negligible and the total entropy is thus dominated by the spin sector~\cite{Lu2026}. This signals a dip in spin entropy, indicating the strongest AFM spin correlations around $U=U_{\rm S}$~\cite{Lu2026}. For larger $U$, these correlations are suppressed by the weakened superexchange coupling $J=4t^2/U$ and the elevated $T/J$ (at fixed $T/t$) within the effective Heisenberg model~\cite{Delannoy2005}. Thus, these insights justify interpreting the extrapolated $U_{\rm cross}$ in Fig.~\ref{fig:Fig01Cross} as the crossover boundary.

We next show how the sign of $(\partial D/\partial T)_U$ near $T=0$ distinguishes between Slater and Mott physics, based on ED analysis on a $4\times 4$ system with PBC. The double occupancy is given by $D(T) = \sum_\ell D_\ell e^{-\beta E_\ell} / \sum_{\ell^{\prime}} e^{-\beta E_{\ell^{\prime}}}$ with $\beta=1/k_BT$, where $E_\ell$ and $|\phi_\ell\rangle$ are the energy and wavefunction of the $\ell$-th eigenstate (and $\ell=0$ labels the ground state), and $D_\ell=\langle \phi_\ell|N_s^{-1}\sum_{\mathbf{i}}\hat{n}_{\mathbf{i}\uparrow}\hat{n}_{\mathbf{i}\downarrow}|\phi_\ell\rangle$. We find that all degenerate eigenstates involved in our ED calculations share the same $D_\ell$, and that the ground state is nondegenerate for any $U>0$. For simplicity, we denote $g_m$ as the degeneracy of the $m$-th excited state ($m\ge1$), with the eigenenergy $E_m$. At low temperatures, the summations over $\ell$ and $\ell^{\prime}$ in $D(T)$ can be truncated when $\beta (E_{M+1} - E_0) \gg 1$, as the higher excited states are thermally suppressed, yielding~\cite{Suppl}
\begin{equation}\begin{aligned}
D(T) \approx D_0 + \frac{\sum_{m=1}^M g_m (D_m - D_0) e^{-\beta(E_m - E_0)}}{1 + \sum_{m^{\prime}=1}^M g_{m^{\prime}} e^{-\beta(E_{m^{\prime}} - E_0)}},
\label{eq:truncatedBZ}
\end{aligned}\end{equation}
where $D_m$ and $D_0$ are the double occupancy of the $m$-th excited state and the ground state, respectively. This equation illustrates that, provided $D_m>D_0$ holds for all involved excited states ($1\le m\le M$), then one obtains $D(T)>D_0$ and $(\partial D/\partial T)_U>0$ (and conversely, $D_m<D_0$ leads to $D(T)<D_0$ and $(\partial D/\partial T)_U<0$)~\cite{Suppl}. It is more evident from the approximation of Eq.~(\ref{eq:truncatedBZ}) near $T=0$ as $D(T)\approx D_0 + \sum_{m=1}^M g_m (D_m - D_0) e^{-\beta(E_m - E_0)}$~\cite{Suppl}. 

\begin{figure}[t]
\centering
\includegraphics[width=0.950\columnwidth]{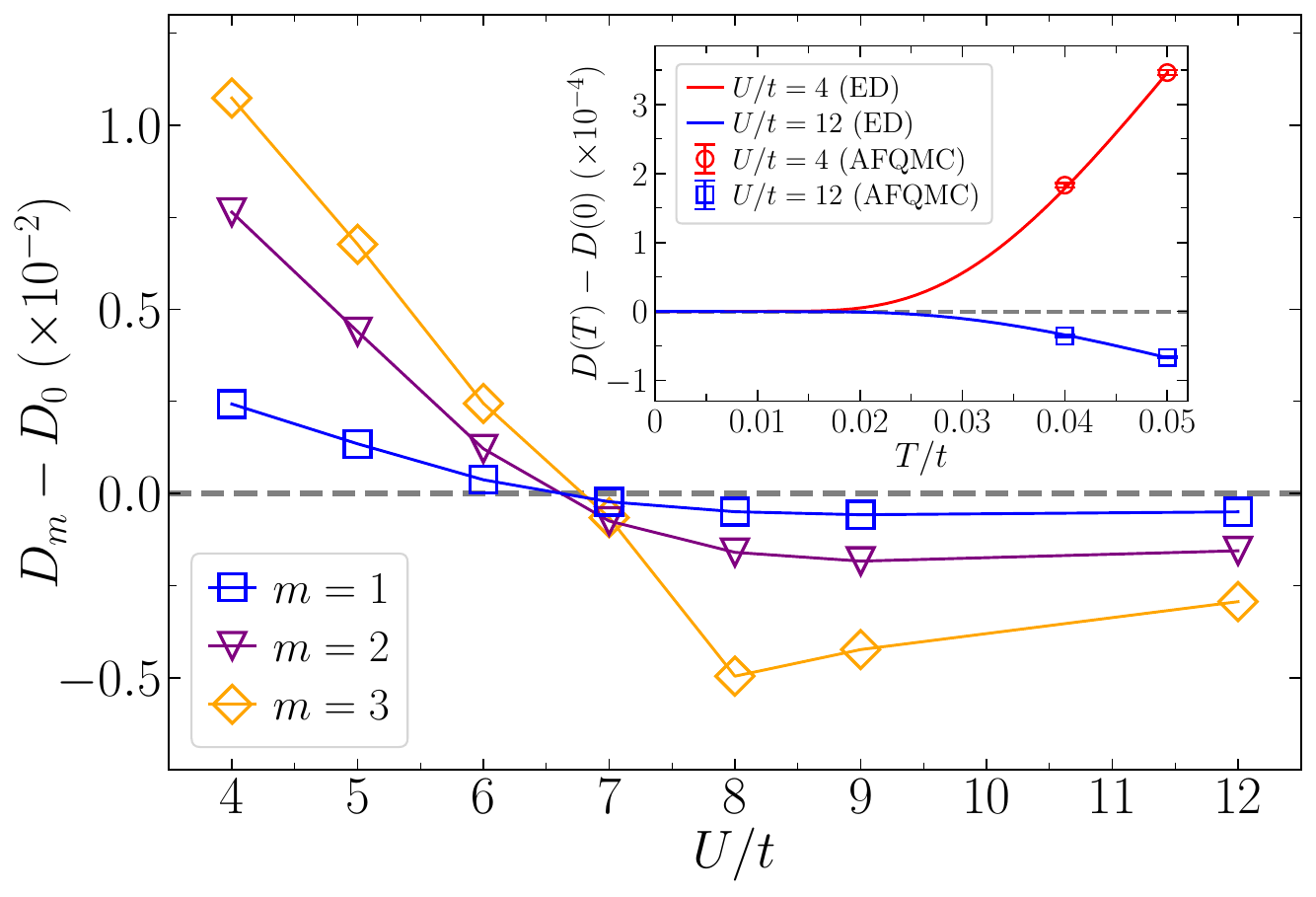}
\caption{
ED results of $(D_m-D_0)$ as a function of $U/t$ with $m=1,2,3$ on a $4\times 4$ periodic supercell for the model~(\ref{eq:Hubbard}), where $D_m$ and $D_0$ are the double occupancy of the $m$-th excited state and the ground state, respectively. The inset plots the difference $[D(T)-D(T=0)]$ versus $T/t$ with $T/t\le0.05$, from both ED [solid lines, with $D(T)$ computed via Eq.~(\ref{eq:truncatedBZ})] and AFQMC (circles and squares) for $U/t=4$ and $12$, with $D(T=0)=D_0$ taken from ED.
}
\label{fig:Fig02ED}
\end{figure}

In ED, we verify that for $T/t\le 0.05$, $M=3$ in Eq.~(\ref{eq:truncatedBZ}) is sufficient to produce accurate results of $D(T)$ across $4\le U/t\le 12$ in the model~(\ref{eq:Hubbard}). As depicted in Fig.~\ref{fig:Fig02ED}, the $(D_m-D_0)$ results clearly demonstrate the aforementioned behaviors of $D_m>D_0$ for $U/t\le6$ and $D_m<D_0$ for $U/t\ge7$. In the small-$U$ regime, the observed hierarchy $D_m>D_{m-1}>\cdots>D_0$ reveals that, with increasing $m$, the low-lying excited states possess enhanced charge redistribution compared to the ground state. Since $D$ directly quantifies the charge fluctuations, these results indicate that the low-lying excitations contributing to the thermal variation of $D$ [i.e., $D(T)$] have a predominantly charge-like character. This characterization does not imply the absence of spin excitations in the Slater regime. Instead, the N\'{e}el AFM order in this regime also supports collective spin excitations, including transverse spin-wave and longitudinal spin-amplitude fluctuations, which are generally intertwined with charge fluctuations, as discussed in Ref.~\cite{Schrieffer1989}. By contrast, in the large-$U$ regime, $D_m<D_0$ is characteristic of spin-like excitations. Here, charge fluctuations become strongly suppressed, and $D_0\propto (t/U)^2$ at ground state~\cite{Avella2003} arises from the virtual hopping~\cite{Delannoy2005}. In low-lying excited states, spin-flip processes create ferromagnetic bonds (in Fock-state basis), which suppress such virtual hopping and thus reduce double occupancy, leading to $D_m<D_0$. Higher spin excitations typically involve more spin flips, rendering $D_m<D_{m-1}<\cdots<D_0$, as indeed observed in Fig.~\ref{fig:Fig02ED}. These understandings are consistent with low-energy spin superexchange physics in the Mott regime.

The results in Fig.~\ref{fig:Fig02ED}, combined with Eq.~(\ref{eq:truncatedBZ}), illustrate that $(\partial D/\partial T)_U>0$ for $U/t\le6$, and $(\partial D/\partial T)_U<0$ for $U/t\ge7$. To further examine these behaviors, we compute $D(T)$ via Eq.~(\ref{eq:truncatedBZ}) using $(g_m,E_m,D_m)$ from ED (see SM~\cite{Suppl}) and compare with finite-$T$ AFQMC results for $U/t=4$ and $12$, as shown in Fig.~\ref{fig:Fig02ED} inset. The nice agreement between the two methods validates the ED results of $D(T)$. As expected, the $D(T)$ curves evidently confirm the predicted sign of $(\partial D/\partial T)_U$, i.e., positive for $U/t=4$ and negative for $U/t=12$. Together, these results and the analyses above establish the connection between the sign of $(\partial D/\partial T)_U$ and the Slater versus Mott regimes via the distinct properties of low-lying excited states. They also provide a direct explanation for the Pomeranchuk effect of $D(T)$ in the Mott regime~\cite{Gorelik2012,GangLi2014,Wietek2021,Qiaoyi2023,Song2025L,*Song2025B,Lu2026}. Quantitatively, the ED results for the $L=4$ system in Fig.~\ref{fig:Fig02ED} locate the Slater-Mott crossover at $U/t=6\sim7$, which is substantially larger than $U_{\rm cross}/t=4.0(2)$ found in Fig.~\ref{fig:Fig01Cross}, a discrepancy we confirm as arising from finite-size effects (see SM~\cite{Suppl}).

\begin{figure}[t]
\centering
\includegraphics[width=0.892\columnwidth]{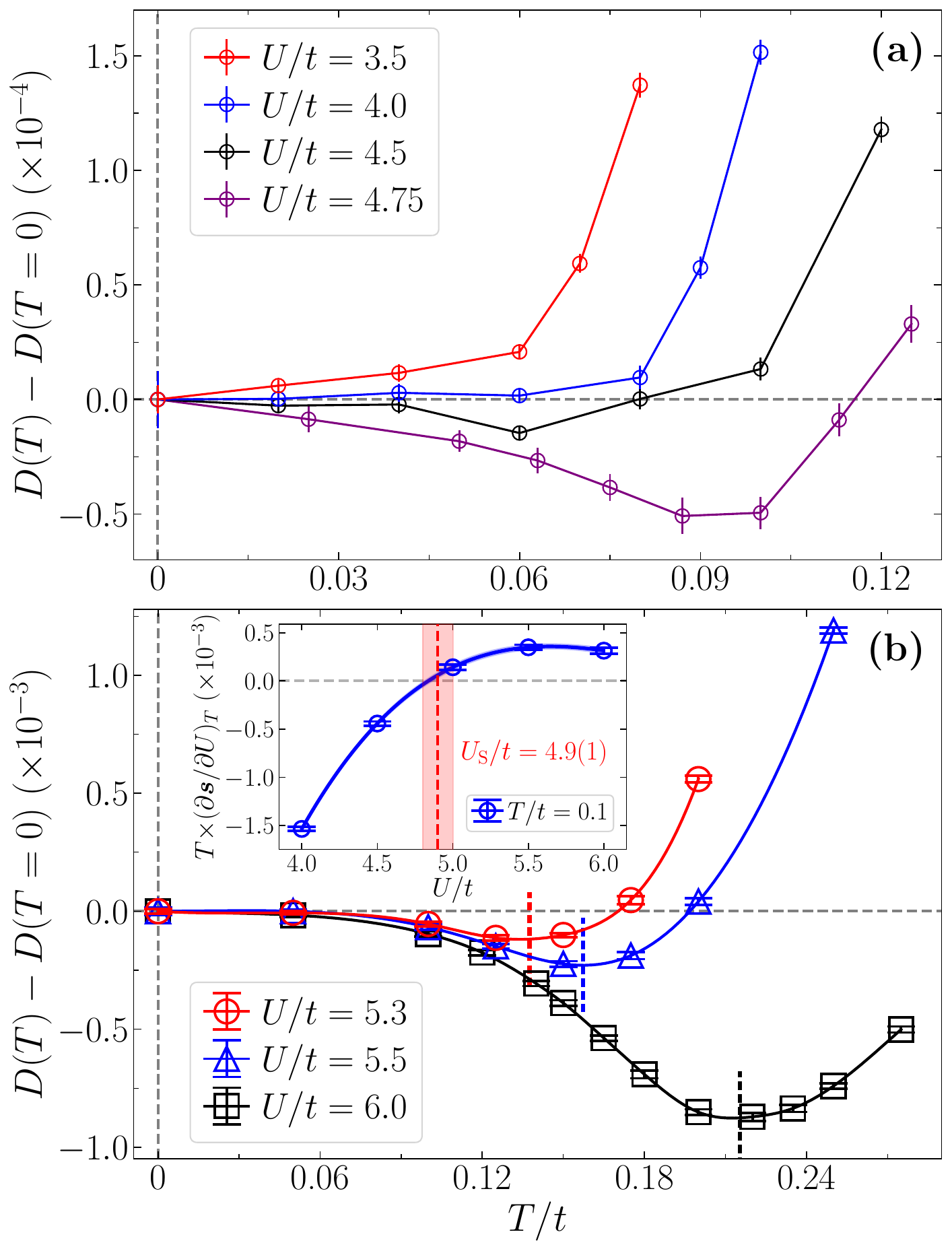}
\caption{
AFQMC results of $[D(T)-D(T=0)]$ as a function of $T/t$ obtained under TABC to reveal the Slater-Mott crossover. Panel (a) presents the results around the crossover, and (b) shows those in the Mott regime, with the solid lines as polynomial fits to the data. In (b), the vertical dashed lines mark the local minimum of $D(T)$ at $T_{\rm D}/t=0.138(4)$, $0.157(3)$, and $0.215(4)$ for $U/t=5.3$, $5.5$, and $6.0$, respectively. The inset of (b) plots $T\times(\partial\boldsymbol{s}/\partial U)_T$ versus $U/t$ at $T/t=0.1$ (see text). The $T=0$ and $T>0$ results are accordingly computed using ground-state and finite-temperature AFQMC with $L=20$, and exhibit negligible finite-size effects.
}
\label{fig:Fig03QMCdata}
\end{figure}

To access larger systems and accurately determine the low-$T$ behavior of $D(T)$, we turn to large-scale AFQMC with TABC. As indicated by $[D(T)-D(T=0)]$ results in Fig.~\ref{fig:Fig03QMCdata}(a), $D(T)$ increases upon heating for $U/t=3.5$, signaling $(\partial D/\partial T)_U>0$ and the Slater regime. In contrast, for $U/t=4.5$ and $4.75$, $D(T)$ decreases with temperature near $T=0$, rendering $(\partial D/\partial T)_U<0$ and thus indicating the Mott regime. At $U/t=4.0$, $D(T)$ instead remains almost constant within statistical uncertainties from $T=0$ up to $T/t=0.06$. This near-constancy indicates that $U/t=4.0$ lies close to the crossover boundary, in agreement with the $U_{\rm cross}/t$ identified in Fig.~\ref{fig:Fig01Cross}. With increasing $U/t$ deeper into the Mott regime, the $(\partial D/\partial T)_U<0$ behavior near $T=0$ becomes more pronounced, as shown in Fig.~\ref{fig:Fig03QMCdata}(b). Moreover, $D(T)$ develops a local minimum at $T_{\rm D}$, which is linked via Eq.~(\ref{eq:Maxwell}) to the local minimum in $\boldsymbol{s}(U)$ at $U_{\rm S}$. We extract $T_{\rm D}/t=0.138(4)$, $0.157(3)$, and $0.215(4)$ for $U/t=5.3$, $5.5$, and $6.0$, respectively, as presented in Fig.~\ref{fig:Fig01Cross}. Additionally, from Fig.~\ref{fig:Fig03QMCdata}(a), we find $T_{\rm D}/t\simeq0.06$ for $U/t=4.5$ and $T_{\rm D}/t\simeq0.09$ for $U/t=4.75$, both consistent with the red shaded region in Fig.~\ref{fig:Fig01Cross}, thus supporting the validity of the linear fit.

Besides extracting $T_{\rm D}$ from $D(T)$, one can also determine the sign change of $(\partial D/\partial T)_U$, based on Eq.~(\ref{eq:Maxwell}), using the local minimum position $U_{\rm S}$ of the entropy $\boldsymbol{s}(U)$. We evaluate $U_{\rm S}$ using three approaches. The first is to explicitly compute $\boldsymbol{s}(U)$ at fixed $T/t$~\cite{Song2025L,*Song2025B,Song2026} and extract $U_{\rm S}$, as shown in Fig.~\ref{fig:Fig01Cross} inset (a). However, this method becomes inefficient at low $T$ since $\boldsymbol{s}(U)$ vanishes as $T/t\to0$. A second approach is to obtain $U_{\rm S}$ from the condition $(\partial\boldsymbol{s}/\partial U)_T=0$, which is equivalent to $(\partial D/\partial T)_U=0$. The latter can be directly measured in AFQMC using $(\partial D/\partial T)_U=\beta^2(\langle \hat{H}\hat{D}\rangle-\langle \hat{H}\rangle\langle \hat{D}\rangle)$~\cite{Suppl} with the operator $\hat{D}=N_s^{-1}\sum_{\mathbf{i}}\hat{n}_{\mathbf{i}\uparrow}\hat{n}_{\mathbf{i}\downarrow}$. As illustrated in Fig.~\ref{fig:Fig01Cross} inset (b), this yields a $U_{\rm S}$ that closely matches that from the first approach. We also find that such a determination of $U_{\rm S}$ exhibits rather weak finite-size effects~\cite{Suppl}, despite the use of $L=8$ in Fig.~\ref{fig:Fig01Cross} inset (b) compared to $L=20$ in Fig.~\ref{fig:Fig01Cross} inset (a). The third approach instead applies $T\times(\partial\boldsymbol{s}/\partial U)_T=(\partial e/\partial U +1/2)-D(U)$~\cite{Suppl} (with $e(U)$ as total energy density) to solve $(\partial\boldsymbol{s}/\partial U)_T=0$ numerically, as depicted in Fig.~\ref{fig:Fig03QMCdata}(b) inset. 
These techniques are used jointly to generate the $U_{\rm S}$ data points, which, combined with the $T_{\rm D}$ data obtained in Fig.~\ref{fig:Fig03QMCdata}(b), extrapolate to $U_{\rm cross}/t=4.0(2)$ at $T=0$ in Fig.~\ref{fig:Fig01Cross}.

\begin{figure}[t]
\centering
\includegraphics[width=0.97\columnwidth]{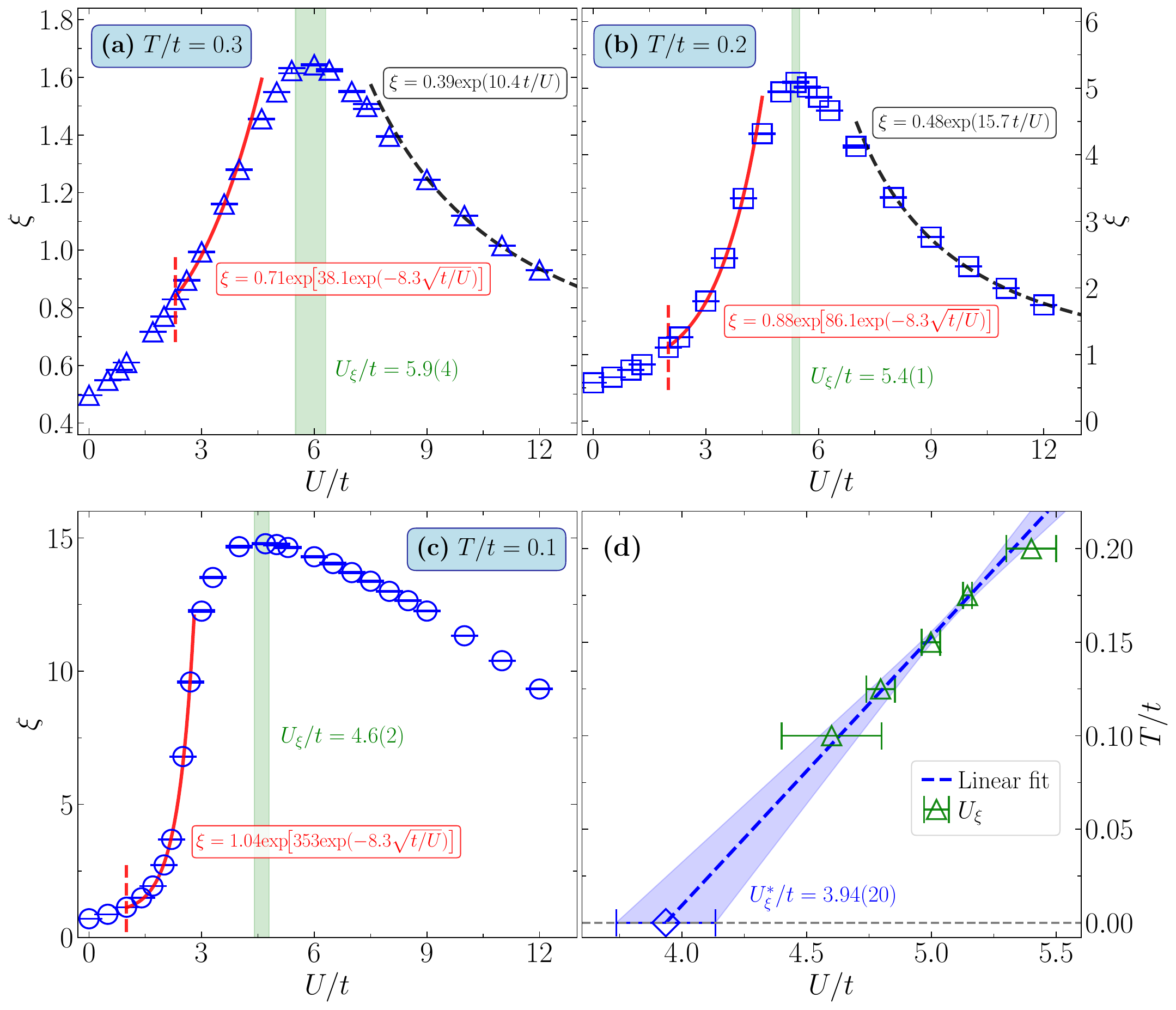}
\caption{AFQMC results for the AFM correlation length $\xi$ obtained under PBC, together with the extrapolation of its maximum position to $T=0$. Panels (a)-(c) present $\xi$ as a function of $U/t$ at $T/t = 0.3$, $0.2$, and $0.1$, respectively, with $L=20$ used in (a)(b) and $L=24$ in (c). The position of the maximum, denoted $U_{\xi}/t$, is indicated by the green shaded band, with the corresponding value included. At weak-coupling side, the $\xi$ data are fitted to $\xi \sim {\rm exp}[A{\rm exp}(-B\sqrt{t/U})]$ (red solid lines) for all three temperatures, with the starting point of the fit marked by the red dashed line. At strong coupling, the data are fitted to $\xi \sim \exp(Ct/U)$ (black dashed lines) for $U/t \ge 9$ at $T/t=0.3$ and $0.2$. Panel (d) shows the extracted $U_{\xi}/t$ at different temperatures and its linear extrapolation for $T/t\le 0.175$, yielding $U_{\xi}^{*}/t=3.94(20)$ at $T=0$, with the blue shaded region indicating the fitting uncertainty.
}
\label{fig:Fig04AFMxi}
\end{figure}

The sign change of $(\partial D/\partial T)_U$, as previously analyzed, is essentially connected to the strongest AFM spin correlation, which also offers key insight into the Slater-Mott crossover. This correlation is dominated by its long-range part at low $T$ and is characterized by the AFM correlation length $\xi$. Practically, $\xi$ can be computed from spin structure factor $S_{zz}(\mathbf{q})=\sum_{\mathbf{r}}e^{{\rm i}\mathbf{q}\cdot\mathbf{r}}C(\mathbf{r})$ with the spin correlation $C(\mathbf{r})=N_s^{-1}\sum_{\mathbf{i}}\langle\hat{s}_{\mathbf{i}}^z\hat{s}_{\mathbf{i}+\mathbf{r}}^z\rangle$ [with $\hat{s}_{\mathbf{i}}^z=(\hat{n}_{\mathbf{i}\uparrow}-\hat{n}_{\mathbf{i}\downarrow})/2$] using $\xi=[2{\rm sin}(\pi/L)]^{-1}\sqrt{S_{zz}(\boldsymbol{\pi})/S_{zz}(\boldsymbol{\pi}+\delta\mathbf{q})-1}$~\cite{Song2025L,*Song2025B}, where the momenta $\boldsymbol{\pi}=(\pi,\pi)$ and $\delta\mathbf{q}=(2\pi/L,0)$ or $(0,2\pi/L)$. In Figs.~\ref{fig:Fig04AFMxi}(a)-(c), we show AFQMC results of $\xi$ versus $U/t$ at $T/t=0.3$, $0.2$, and $0.1$, respectively.

At fixed $T/t$, $\xi$ increases with $U/t$ at weak coupling due to Fermi surface nesting and decreases at strong coupling owing to the reduced $J\propto 1/U$ and the correspondingly elevated $T/J$ within the Heisenberg model. This gives rise to a maximum in $\xi(U)$, with its location denoted as $U_{\xi}$. Upon cooling, the distinct Slater versus Mott mechanisms result in a more pronounced asymmetry of $\xi$ about $U=U_{\xi}$, as observed in Figs.~\ref{fig:Fig04AFMxi}(a)-(c). Moreover, the $U$ dependence of $\xi$ at both sides can be analyzed at a quantitative level. For the 2D model~(\ref{eq:Hubbard}), below a characteristic temperature $T_{X}/t$, $\xi$ is expected to display an exponential form $\xi(T)\sim e^{\alpha/T}$ for any fixed $U/t$~\cite{Thomas2021,Sudip1989,Borejsza2003,Borejsza2004,Beard1998}, with $\alpha$ being a $U$-dependent constant. At weak coupling, $\alpha$ is governed by $\Delta_{sp} \sim t {\rm exp}(-2\pi\sqrt{t/U})$~\cite{Hirsch1985,Edwin2025} as the single-particle gap. It leads to $\xi(U) \sim {\rm exp}[A{\rm exp}(-B\sqrt{t/U})]$ at fixed $T/t$, which is only valid for $U_0\lesssim U\lesssim U_1$ with $U_0$ defined by $T_{X}(U_0)=T$ and $U_1$ marking the breakdown of above asymptotic expression of $\Delta_{sp}$. At $T/t=0.2$, Ref.~\cite{Thomas2021} obtained $U_0/t\simeq2$, which should be smaller (larger) at $T/t=0.1$ ($T/t=0.3$). This is incorporated in the fit of $\xi(U)$ below $U_{\xi}$ (red solid line), as shown in Figs.~\ref{fig:Fig04AFMxi}(a)-(c). 
At strong coupling, $\alpha$ is instead set by $J=4t^2/U$~\cite{Beard1998}, which yields $\xi(U)\sim{\rm exp}(Ct/U)$ at fixed $T/t$. Our $\xi(U)$ data support such scaling for $U/t\ge9$ at $T/t=0.3$ and $0.2$, with the extracted $C$ agreeing well with a $1/T$ form, as seen from the fit (black dashed line) in Figs.~\ref{fig:Fig04AFMxi}(a)-(b). Finite-size effects in $\xi(U)$ are noticeable around $U_{\xi}$ (absent at $T/t=0.3$, small at $0.2$, significant at $0.1$), while $U_{\xi}$ itself converges rapidly for the system sizes used. Notably, at $T/t=0.1$, the maximal $\xi$ ($\sim$$15$) and its values at $9\le U/t\le12$ remain comparable to $L/2$ of the $L=24$ system [see Fig.~\ref{fig:Fig04AFMxi}(c)], indicating that $\xi$ is cut off by the finite system size. Such significant finite-size effects depend on $U/t$, and therefore prevent a meaningful fit to the asymptotic exponential form for the $\xi(U)$ data.

Near $U\sim U_{\xi}$, where $J$ and $T/J$ exhibit little variation, the rapid suppression of $\xi(U)$ once $U>U_{\xi}$ is instead attributed to the high-order frustrating interactions in the effective spin model~\cite{Delannoy2005}. These include the next-nearest-neighbor and ring-exchange couplings, both $\propto 1/U^3$. The long-range AFM spin correlation is highly vulnerable to such frustrations, which causes $\xi$ to decrease. Thus, $U_{\xi}$ at low $T$ signals the onset of spin superexchange. We note that the peak in $\xi(U)$ and the sign change of $(\partial D/\partial T)_U$ (or the entropy minimum at $U_{\rm S}$) are not directly related. Instead, they are two independent thermal manifestations of the same underlying crossover: the emergence of spin-superexchange dominated physics once charge fluctuations are sufficiently suppressed. While the entropy minimum at $U_{\rm S}$ indicates similar physics, we find $U_{\xi}$ is consistently slightly smaller than $U_{\rm S}$ at the same $T/t$~\cite{Suppl}. This is because $U_{\rm S}$ reflects the overall AFM spin correlations and is therefore less sensitive to frustrations than $\xi$ is. This discrepancy gradually vanishes as $T/t\to0$, when long-range AFM correlations dominate. Hence, extrapolating $U_{\xi}$ to $T=0$ should independently determine the Slater-Mott crossover boundary. We perform this procedure in Fig.~\ref{fig:Fig04AFMxi}(d) using a linear fit to the $U_{\xi}/t$ data, reaching $U_{\xi}^{*}/t=3.94(20)$. This value is in nice agreement with $U_{\rm cross}/t=4.0(2)$ from Fig.~\ref{fig:Fig01Cross}, thus providing compelling support for our physical picture of the crossover and the Slater versus Mott mechanisms.

All the above physical interpretations and calculations can be readily applied to a group of Hubbard models hosting the Slater-Mott crossover, including the 2D modified model with spin-nematic hopping~\cite{Gukelberger2017,Xie2025}, the bilayer model~\cite{Kancharla2007,Golor2014,Gall2021}, the SU(2N) models (for instance $N=2$ and $3$)~\cite{Cai2013,Zhou2014,Wang2014,Wang2019}, and the 3D model on simple cubic lattice~\cite{Song2025L,*Song2025B,Song2025CPL}. In particular, for the half-filled 3D Hubbard model, we obtain $U_{\mathrm{cross}}^{\rm 3D}/t=5.7(2)$ as the crossover boundary from the analysis of AFM correlation length (see SM~\cite{Suppl}). Our numerical results also have implications for optical lattice experiments~\cite{Mazurenko2017,Xu2025,Kendrick2025,Bourgund2025,Shao2024,YuXuan2025,Thomas2026}, which have now entered the low-$T$ regime ($T/t\sim0.05$)~\cite{Xu2025} of the Hubbard model and can directly measure both double occupancy~\cite{YuXuan2025} and AFM correlation length~\cite{Mazurenko2017}. Moreover, recent double occupancy measurements have achieved a precision better than $0.1\%$ and serves as a decisive indicator of lattice superfluidity~\cite{CommNote}. Given these capabilities and further improvements, future optical lattice experiments should be able to resolve the $D(T)$ and $\xi(U)$ results in Figs.~\ref{fig:Fig03QMCdata} and~\ref{fig:Fig04AFMxi}, thereby directly visualizing the Slater-Mott crossover.

We then briefly comment on other physical observables that are related to the Slater and Mott physics in the half-filled Hubbard model. The local moment is directly tied to $D$ as $\langle m_z^2\rangle=N_s^{-1}\sum_{\mathbf{i}}\langle(\hat{n}_{\mathbf{i}\uparrow}-\hat{n}_{\mathbf{i}\downarrow})^2/4\rangle=1/4-D/2$~\cite{Thereza2001}, and its low-$T$ derivative would exhibit an opposite sign-change behavior to $(\partial D/\partial T)_U$, thus offering a similar criterion for the Slater-Mott crossover. The charge compressibility $\chi_e$~\cite{Lu2026} is strictly zero at $T=0$ for the insulating ground state, and $(\partial\chi_e/\partial T)_U>0$ near $T=0$ holds for any $U>0$~\cite{Kim2020}, providing no distinction between the two regimes. The uniform spin susceptibility $\chi_s$ exhibits a similar behavior, with $(\partial\chi_s/\partial T)_U>0$ near $T=0$ for any $U>0$~\cite{Kim2020,Moreo1993}. As for the kinetic energy $e_K$, our high-precision AFQMC results explicitly show $(\partial e_K/\partial T)_U>0$ in the low-$T$ regime across the crossover (see SM~\cite{Suppl}), which does not support a clear criterion. These observations reinforce that our proposed criterion based on the sign change of $(\partial D/\partial T)_U$ is a direct, highly physically relevant, and experimentally accessible marker of the Slater-Mott crossover.

In summary, we have established a simple and quantitative thermal criterion for the Slater-Mott crossover in the Hubbard model in terms of the temperature dependence of double occupancy near $T=0$, i.e., $(\partial D/\partial T)_U>0$ in the Slater regime and $(\partial D/\partial T)_U<0$ in the Mott regime. Using ED calculations, we validate this criterion by linking the sign of $(\partial D/\partial T)_U$ to the low-lying excited state properties characteristic of Slater and Mott physics. We further verify the criterion via low-temperature behavior of $D(T)$ obtained from large-scale AFQMC simulations, and pinpoint $U_{\mathrm{cross}}/t=4.0(2)$ as the crossover boundary for the 2D Hubbard model. Moreover, through the Maxwell relation, the sign change of $(\partial D/\partial T)_U$ is connected to the entropy minimum and, consequently, to the maximum of AFM correlation length, from which we reach a consistent boundary estimate. Beyond locating the crossover, our work offers a direct thermal perspective on the ground-state Slater-Mott crossover that is readily accessible in optical lattice experiments.

\begin{acknowledgments}
{\it Acknowledgments}. This work was supported by the National Natural Science Foundation of China (Grants No. 12247103, No. 12204377, and No. 12275263), the Quantum Science and Technology-National Science and Technology Major Project (Grant No. 2021ZD0301900), the Natural Science Foundation of Fujian province of China (Grant No. 2023J02032), and the Youth Innovation Team of Shaanxi Universities.
\end{acknowledgments}

\bibliography{SlaterToMottRef.bib}


\onecolumngrid
\newpage

\startsupplement

\begin{center}
\textbf{\large Supplementary Material for \\
``Thermal Signatures of the Slater-Mott Crossover in the Hubbard model: \\ From Double Occupancy to Antiferromagnetic Correlation Length''}
\end{center}

\section{I. Overview of this supplementary material}
\label{sec:Overview}

This Supplemental Material is divided into two parts. First, we concentrate on the details of the auxiliary-field quantum Monte Carlo (AFQMC) simulations applied in this work, which include the key algorithmic ingredients, the computation of thermal entropy $\boldsymbol{s}(U)$ and its local minimum $U_{\rm S}$, and the table summarizing the signal locations of $U_{\rm S}$, $T_{\rm D}$, and $U_{\xi}$. Second, we provide the proof for Eq.~(3) in the main text, the tables listing ED results of $(g_m,E_m,D_m)$ used to obtain the $D(T)$ results in Fig.~2 inset in the main text, and additional results from ED calculations regarding the sign of $(\partial D/\partial T)_U$.

\section{II. AFQMC algorithm and calculations for physical observables}
\label{sec:AFQMCPhyObs}

\subsection{A. Key ingredients of the AFQMC algorithm}
\label{sec:AFQMC}

In this work, we employ both the finite-temperature and ground-state AFQMC methods, which are free of the minus sign problem for the half-filled Hubbard model we study. Here, we briefly introduce the methods and summarize the key ingredients applied in our simulations. 

The core principle of the AFQMC algorithm~\cite{Blankenbecler1981,Hirsch1983,White1989,Scalettar1991,Congjun2005,Assaad2008,He2019B,Sun2024,Song2025L,*Song2025B,Yuanyao2025,Sugiyama1986,Sorella1989,Duhao2025} is to decompose the two-body interaction into noninteracting fermions coupled to auxiliary fields using Hubbard-Stratonovich (HS) transformations~\cite{Hirsch1983}, and then to compute the fermionic observables through importance sampling of these auxiliary fields. Among all ingredients of AFQMC algorithm, the HS transformation plays a central role, as it is directly linked to both the sign problem and the statistical uncertainty of specific observables. 

The finite-temperature AFQMC method~\cite{Blankenbecler1981,Hirsch1983,White1989,Scalettar1991,Assaad2008,He2019B,Sun2024,Song2025L,*Song2025B,Yuanyao2025,Sugiyama1986,Sorella1989,Duhao2025} starts from the imaginary-time discretization as $\beta=M\Delta\tau$, and expresses the partition function as $Z=\text{Tr}(e^{-\beta\hat{H}})=\text{Tr}[(e^{-\Delta\tau\hat{H}})^M]$. Then the symmetric Trotter-Suzuki decomposition as $e^{-\Delta\tau\hat{H}}=e^{-\Delta\tau\hat{H}_0/2}e^{-\Delta\tau\hat{H}_I}e^{-\Delta\tau\hat{H}_0/2}+\mathcal{O}[(\Delta\tau)^3]$ is applied to separate the interaction term $\hat{H}_I$ from free fermion part $\hat{H}_0$. The Trotter error $\mathcal{O}[(\Delta\tau)^3]$ is typically eliminated by the extrapolation to $\Delta\tau\to0$ limit. The calculation of a physical observable applies the expression $\langle\hat{O}\rangle=\text{Tr}(e^{-\beta\hat{H}}\hat{O})/\text{Tr}(e^{-\beta\hat{H}})$. The ground-state AFQMC method~\cite{White1989,Assaad2008,Sugiyama1986,Sorella1989,Duhao2025} obtains the ground-state wavefunction $|\Psi_g\rangle$ via imaginary-time projection from an initial wavefunction $|\psi_T\rangle$ as $|\Psi_g\rangle=\lim_{\Theta\to\infty}e^{-\Theta\hat{H}}|\psi_T\rangle$, where $\Theta$ is the projection parameter. The above imaginary-time discretization for $\Theta$ and Trotter-Suzuki decomposition for $e^{-\Delta\tau\hat{H}}$ is similarly applied, and the evaluation of a physical observable relies on $\langle\hat{O}\rangle=\langle\Psi_g|\hat{O}|\Psi_g\rangle/\langle\Psi_g|\Psi_g\rangle=\lim_{\Theta\to\infty}\langle\psi_T|e^{-\Theta\hat{H}/2}\hat{O}e^{-\Theta\hat{H}/2}|\psi_T\rangle/\langle\psi_T|e^{-\Theta\hat{H}}|\psi_T\rangle$. 

In our AFQMC simulations, we combine the HS transformation into the spin-$\hat{s}^z$ (HS-$\hat{s}^z$) channel and into the charge-density (HS-$\hat{n}$) channel~\cite{Hirsch1983,Song2025B} to achieve high-precision results. The HS-$\hat{s}^z$ transformation reads
\begin{equation}\begin{aligned}
\label{eq:HSspinDecomp}
e^{-\Delta\tau U \big(\hat{n}_{\mathbf{i}\uparrow} \hat{n}_{\mathbf{i}\downarrow} - \frac{\hat{n}_{\mathbf{i}\uparrow} + \hat{n}_{\mathbf{i}\downarrow}}{2}\big) }
= C_s\sum_{x_{\mathbf{i}}=\pm1}e^{\gamma_s x_{\mathbf{i}}(\hat{n}_{\mathbf{i}\uparrow}-\hat{n}_{\mathbf{i}\downarrow})}
\end{aligned}\end{equation}
with the constant $C_s=1/2$, and the coupling coefficient $\gamma_s=\cosh^{-1}(e^{+\Delta\tau U/2})$ for $U>0$ and $\gamma_s=i\cos^{-1}(e^{+\Delta\tau U/2})$ for $U<0$. The HS-$\hat{n}$ transformation is given as
\begin{equation}\begin{aligned}
\label{eq:HScharge}
e^{-\Delta\tau U \big(\hat{n}_{\mathbf{i}\uparrow} \hat{n}_{\mathbf{i}\downarrow} - \frac{\hat{n}_{\mathbf{i}\uparrow} + \hat{n}_{\mathbf{i}\downarrow}}{2}\big) }
= C_c \sum_{x_{\mathbf{i}} = \pm 1} e^{\gamma_c x_{\mathbf{i}} (\hat{n}_{\mathbf{i}\uparrow} + \hat{n}_{\mathbf{i}\downarrow} - 1)},
\end{aligned}\end{equation}
with the constant $C_c=e^{+\Delta\tau U/2}/2$, and the coupling coefficient $\gamma_c=i\cos^{-1}(e^{-\Delta\tau U/2})$ for $U>0$ and $\gamma_c=\cosh^{-1}(e^{-\Delta\tau U/2})$ for $U<0$. Previous studies~\cite{Song2025B,Xie2025} showed that this mixed-channel implementation of the HS transformations can substantially suppress statistical fluctuations of measured observables in AFQMC. Specifically, we employ HS-$\hat{s}^z$ to compute density-related quantities, such as the double occupancy, total energy, and thermal entropy (Fig.~1 insets, Fig.~2, and Fig.~3 in the main text). Instead, we use HS-$\hat{n}$ to evaluate spin-related properties, including the spin-spin correlations and AFM correlation length (Fig.~4 in the main text). Several advanced techniques are further integrated into our AFQMC algorithm to enhance the simulation efficiency. These include the fast Fourier transform~\cite{Yuanyao2025} and the delayed update technique~\cite{Duhao2025,Sun2024}, both extended to ground-state and finite-temperature AFQMC methods. In the finite-temperature simulations, we have also applied the $\tau$-line global update~\cite{Scalettar1991}. Together, these ingredients enable highly efficient simulations of the Hubbard model in this work.

In practical calculations, we adopt $\Delta\tau t=0.02$ and verify that the residual Trotter error is well below the statistical error in all our numerical results, rendering it negligible. This is particularly crucial for double occupancy, whose high precision makes it especially susceptible to significant Trotter errors~\cite{Song2026}. We apply periodic and twist-averaged boundary conditions (PBC and TABC) to compute the AFM correlation length (Fig.~4 of the main text) and the double occupancy (and the derivatives $(\partial D/\partial T)_U$ and $(\partial\boldsymbol{s}/\partial U)_T$, Fig.~3 of the main text), respectively. 

\subsection{B. Computation of thermal entropy \texorpdfstring{$\boldsymbol{s}(U)$}{} and its local minimum \texorpdfstring{$U_{\rm S}$}{}}
\label{sec:ComputeEntropy}

In this work, the computation of thermal entropy as a function of $U$ at fixed $T/t$, denoted as $\boldsymbol{s}(U)$, plays a significant role in identifying the Slater-Mott crossover. Besides, we have applied three different methods to determine the local minimum position $U_{\rm S}$ of $\boldsymbol{s}(U)$. 

We evaluate $\boldsymbol{s}(U)$ in the half-filled Hubbard model via the following formula established in Ref.~\cite{Song2025L,*Song2025B,Song2026} as
\begin{equation}\begin{aligned}
\label{eq:Entropy}
\boldsymbol{s}(U)=\frac{1}{T}\Big[e(U)-\Omega_0-\int_0^U D(U^{\prime})dU^{\prime}+U/2\Big],
\end{aligned}\end{equation}
where $e(U)=\langle\hat{H}\rangle/N_s$ is the total energy density and $\Omega_0=-2(T/N_s)\sum_{\mathbf{k}}{\rm ln}(1+e^{-\beta \varepsilon_{\mathbf{k}}})$ (with $\varepsilon_{\mathbf{k}}$ as the kinetic energy dispersion) is the grand potential density at $U=0$. This formula only involves the total energy and double occupancy, which can be easily accessed by various many-body numerical methods. 

\begin{figure}[t]
\centering
\includegraphics[width=0.8\textwidth]{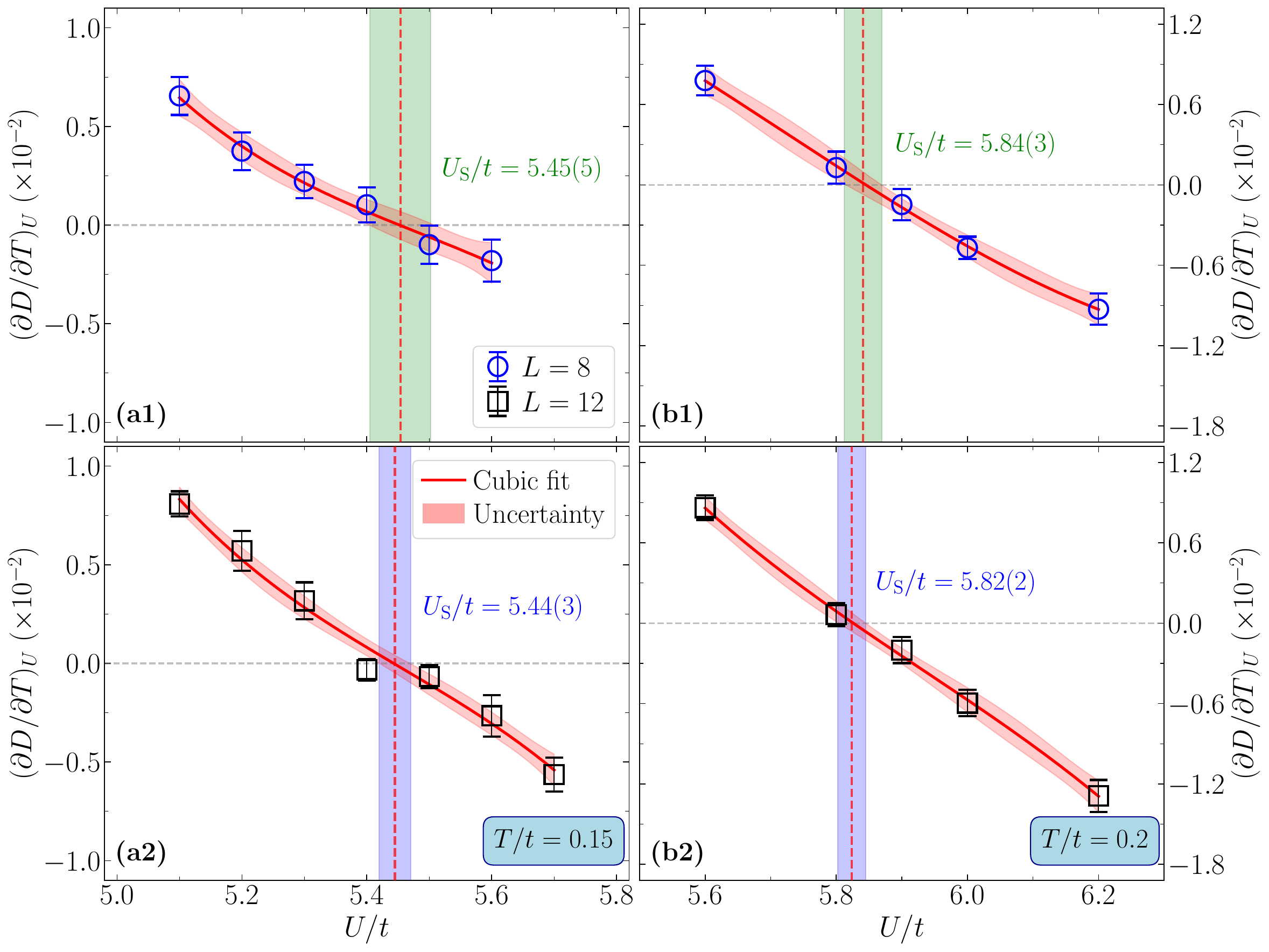}
\caption{
AFQMC results for $(\partial D/\partial T)_U$ as a function of $U/t$ under TABC for (a1)(a2) $T/t=0.15$ and (b1)(b2) $T/t=0.20$. Panels (a1) and (b1) show the results from $L=8$, while (a2) and (b2) are for $L=12$. The vertical dashed line marks the extracted $U_{\rm S}$ from the condition $(\partial D/\partial T)_U=0$, with the shaded region indicating the fitting uncertainty.
}
\label{fig:FigS1pDpT}
\end{figure}

We next discuss the three different methods to compute $U_{\rm S}$. {\it First}, we explicitly evaluate $\boldsymbol{s}(U)$ via Eq.~(\ref{eq:Entropy}), and then directly extract the value of $U_{\rm S}$, which is illustrated in Fig.~1 inset (a) in the main text. This method tends to be more difficult and inefficient at low temperatures, since $\boldsymbol{s}(U)$ gradually vanishes as $T/t$ decreases. {\it Second}, based on Eq.~(\ref{eq:Entropy}), we can directly compute the following derivative 
\begin{equation}\begin{aligned}
\label{eq:PsPu}
\Big(\frac{\partial\boldsymbol{s}}{\partial U}\Big)_T = \frac{1}{T}\big[(\partial e/\partial U +1/2)-D(U)\big],
\end{aligned}\end{equation}
and obtain $U_{\rm S}$ via the condition $(\partial\boldsymbol{s}/\partial U)_T=0$. In Fig.~3(b) inset in the main text, we show the results $T\times(\partial\boldsymbol{s}/\partial U)_T$ at $T/t=0.1$ as an illustration for this method. This method can be applied in the low-temperature regime. {\it Third}, we can transform the condition $(\partial\boldsymbol{s}/\partial U)_T=0$ to $(\partial D/\partial T)_U=0$ using the Maxwell relation, and then directly measure the derivative $(\partial D/\partial T)_U$ versus $U/t$ in AFQMC simulations to determine the value of $U_{\rm S}$. The double occupancy $D(T)$ can be expressed as
\begin{equation}\begin{aligned}
D(T)\equiv \langle \hat D\rangle = \frac{\mathrm{Tr}(e^{-\beta \hat H}\hat D)}{Z},
\end{aligned}\end{equation}
with $Z=\mathrm{Tr}(e^{-\beta \hat H})$ as the partition function and $\hat{D}=N_s^{-1}\sum_{\mathbf{i}}\hat{n}_{\mathbf{i}\uparrow}\hat{n}_{\mathbf{i}\downarrow}$ as the double occupancy operator. Then we can compute the derivative 
\begin{equation}\begin{aligned}
\label{eq:pDpT}
\Big(\frac{\partial D}{\partial T}\Big)_U
= \frac{\partial D}{\partial \beta}\times \frac{\partial \beta}{\partial T}
= -\beta^2 \frac{\partial}{\partial \beta}\Big(\frac{\mathrm{Tr}(e^{-\beta \hat H}\hat D)}{Z}\Big).
\end{aligned}\end{equation}
Applying the following equalities
\begin{equation}\begin{aligned}
\frac{\partial}{\partial \beta}\mathrm{Tr}(e^{-\beta \hat H}\hat D)=-\mathrm{Tr}(e^{-\beta \hat H}\hat H\hat D), \qquad \frac{\partial Z}{\partial \beta}=-\mathrm{Tr}(e^{-\beta \hat H}\hat H),
\end{aligned}\end{equation}
we can simplify Eq.~(\ref{eq:pDpT}) as
\begin{equation}\begin{aligned}
\label{eq:pDpT2}
\Big(\frac{\partial D}{\partial T}\Big)_U
= \beta^2\big(\langle \hat H\hat D\rangle-\langle \hat H\rangle\langle \hat D\rangle\big).
\end{aligned}\end{equation}
Via this formula, we measure the correlation $(\langle \hat H\hat D\rangle-\langle \hat H\rangle\langle \hat D\rangle)$ in AFQMC calculations and compute $(\partial D/\partial T)_U$ as a function of $U/t$ at fixed $T/t$, and extract $U_{\rm S}$ from the condition $(\partial D/\partial T)_U=0$. In Fig.~1 inset (b) in the main text, we show the demonstration results of $(\partial D/\partial T)_U$ for $T/t=0.2$, from which we obtain a consistent value of $U_{\rm S}$ with that in Fig.~1 inset (a), thus validating this method for calculating $U_{\rm S}$. Furthermore, we find that the $U_{\rm S}$ result obtained via this method shows rather weak finite-size effects. As shown in Fig.~\ref{fig:FigS1pDpT}, we compare the AFQMC results (using TABC) of $(\partial D/\partial T)_U$ from $L=8$ and $L=12$ at $T/t=0.15$ [(a1)(a2)] and $T/t=0.20$ [(b1)(b2)]. For both temperatures, the $U_{\rm S}$ values extracted from $L=8$ and $L=12$ are well consistent within the fitting uncertainties, which indicates the negligible finite-size effects already for $L=8$. 

\subsection{C. The signal locations of \texorpdfstring{$U_{\rm S}$}{}, \texorpdfstring{$T_{\rm D}$}{}, and \texorpdfstring{$U_{\xi}$}{}}
\label{sec:SignalLocation}

In Table~\ref{Tab01SignalLocations}, we list the signal locations obtained from AFQMC calculations, including the local minimum $U_{\rm S}$ of thermal entropy $\boldsymbol{s}(U)$, the maximum $U_{\xi}$ of AFM correlation length $\xi(U)$, and the local minimum $T_{\rm D}$ of double occupancy $D(T)$ in the Mott regime ($U>U_{\rm cross}$). These signal locations are verified to converge to the thermodynamic limit. The $U_{\rm S}$ and $T_{\rm D}$ results are shown in Fig.~1 in the main text, while the $U_{\xi}$ data are included in Fig.~4(d).

\begin{table}[h]
\caption{\label{Tab01SignalLocations}
Signal locations of $U_{\rm S}/t$, $T_{\rm D}/t$, and $U_{\xi}/t$ shown in Fig.~1 and Fig.~4(d) in the main text. The first three columns list the results from fixed-temperature scans, while the last two columns present results from fixed-$U$ calculations. 
}
\centering
\setlength{\tabcolsep}{6pt}
\begin{tabular}{l l l|l l}
\hline\hline
\multicolumn{1}{c}{$T/t$} & \multicolumn{1}{c}{$U_{\mathrm{S}}/t$} & \multicolumn{1}{c|}{$U_{\xi}/t$} & \multicolumn{1}{c}{$U/t$} & \multicolumn{1}{c}{$T_{\mathrm{D}}/t$} \\
\hline
0.20 & 5.9(1)  &        & 5.5 & 0.138(4) \\
0.175 & 5.63(3) & 5.14(2) & 5.5 & 0.157(3) \\
0.15 & 5.44(3) & 5.00(4) & 6.0 & 0.215(4) \\
0.125 &        & 4.79(6) &   &          \\
0.10 & 4.9(1)  &        &   &          \\
\hline\hline
\end{tabular}
\end{table}

\section{III. Proof for Eq.~(3) in the main text and additional ED results}
\label{sec:EDdouOcc}

In this subsection, we first derive Eq.~(3) and then present additional numerical results from ED for Fig.~3 in the main text. We also discuss the sign of $(\partial D/\partial T)_U$ based on these formulas.

Starting from a general observable $\hat{O}$, its grand-canonical ensemble average at a finite temperature $T$ (with $\beta=1/T$), denoted as $\langle\hat{O}\rangle(T)$, can be expressed in the Lehmann representation as
\begin{equation}\begin{aligned}
\label{eq:ThermalAvg0}
\langle O \rangle(T)
=\frac{{\rm Tr}(\hat{O}e^{-\beta\hat{H}})}{{\rm Tr}(e^{-\beta\hat{H}})}
=\frac{\sum_{\ell}O_{\ell}e^{-\beta E_{\ell}}}{\sum_{\ell^{\prime}}e^{-\beta E_{\ell^{\prime}}}},
\end{aligned}\end{equation}
where $E_\ell$ and $|\phi_\ell\rangle$ are the energy and wavefunction of the $\ell$-th eigenstate (and $\ell=0$ labels the ground state), and $O_\ell=\langle \phi_\ell|\hat{O}|\phi_\ell\rangle$. Practically, the many-body energy levels $E_\ell$ can have degeneracy for the Hubbard model we study. Thus, we introduce $g_m$ as the degeneracy of the $m$-th excited state (with $m\ge 1$), and the corresponding energy level spectrum $\{E_0,E_m\}$ satisfies $E_0<E_1<E_2<\cdots$. In our ED calculations, we find that the ground state of the half-filled 2D Hubbard model in Eq.~(1) in the main text is nondegenerate for any $U>0$ (or equivalently $g_0=1$). Then Eq.~(\ref{eq:ThermalAvg0}) can be rewritten as
\begin{equation}\begin{aligned}
\label{eq:ThermalAvg1}
\langle O\rangle(T)
= \frac{O_0 e^{-\beta E_0} + \sum_{m\ge1}g_m O_{m}e^{-\beta E_{m}}}{e^{-\beta E_0} + \sum_{m^{\prime}\ge1}g_{m^{\prime}} e^{-\beta E_{m^{\prime}}}}
= O_0 + \frac{\sum_{m\ge1} g_m (O_m - O_0) e^{-\beta(E_m - E_0)}}{1 + \sum_{m^{\prime}\ge1} g_{m^{\prime}} e^{-\beta(E_{m^{\prime}} - E_0)}}.
\end{aligned}\end{equation}
Here we assume that for all the $g_m$ degenerate states, $O_m$ share the same value. This is indeed the case for the double occupancy. If this condition does not hold, the quantity $O_m$ can be instead computed as $O_m=g_m^{-1}\sum_{\ell=1}^{g_m}\langle\phi_{m,\ell}|\hat{O}|\phi_{m,\ell}\rangle$, with $|\phi_{m,\ell}\rangle$ denoting the $\ell$-th state wavefunction for the $m$-th excited state. At low temperatures, the contributions of high-energy states are suppressed by the Boltzmann factor $e^{-\beta(E_m-E_0)}$. It is therefore natural to truncate the sum over $m$ and $m^{\prime}$ to the first $M$ excited states for which this factor is still not negligible. This gives the truncated expression
\begin{equation}\begin{aligned}
\label{eq:ThermalAvgTruc}
\langle O\rangle_{\rm M}(T) \approx O_0 + \frac{\sum_{m=1}^M g_m (O_m - O_0) e^{-\beta(E_m - E_0)}}{1 + \sum_{m^{\prime}=1}^M g_{m^{\prime}} e^{-\beta(E_{m^{\prime}} - E_0)}},
\end{aligned}\end{equation}
which holds under the condition $\beta (E_{M+1} - E_0) \gg 1$ (or $e^{-\beta(E_{M+1} - E_0)}\ll 1$). Taking $\hat{O}=\hat{D}=N_s^{-1}\sum_{\mathbf{i}}\hat{n}_{\mathbf{i}\uparrow}\hat{n}_{\mathbf{i}\downarrow}$, we can reach the following expression
\begin{equation}\begin{aligned}
\label{eq:TrucForD}
\langle D\rangle_{\rm M}(T) \approx D_0 + \frac{\sum_{m=1}^M g_m (D_m - D_0) e^{-\beta(E_m - E_0)}}{1 + \sum_{m^{\prime}=1}^M g_{m^{\prime}} e^{-\beta(E_{m^{\prime}} - E_0)}},
\end{aligned}\end{equation}
which is exactly the Eq.~(3) in the main text. 

\begin{table}[h]
\caption{\label{Tab02EDLevels}
ED results of $(E_m,D_m,g_m)$ for the ground state and the lowest three excited levels on a $4\times4$ periodic system. The left and right tables correspond to $U/t=4$ and $U/t=12$, respectively. Note that the energy $E_m$ corresponds to the many-body eigenenergy level of the model Hamiltonian in the main text [see Eq.~(1)], which contains the term $-U/2\sum_{\mathbf{i}}(\hat{n}_{\mathbf{i}\uparrow} + \hat{n}_{\mathbf{i}\downarrow})$.}
\centering
\begin{minipage}{0.47\textwidth}
\centering
\setlength{\tabcolsep}{4pt}
\begin{tabular}{l l l l}
\hline\hline
\multicolumn{4}{c}{$U/t=4$} \\
\hline
\multicolumn{1}{c}{$m$} & \multicolumn{1}{c}{$E_m/t$} & \multicolumn{1}{c}{$D_m$} & \multicolumn{1}{c}{$g_m$} \\
\hline
0 & -45.621854821163 & 0.115125560655 & 1 \\
1 & -45.476275850419 & 0.117556096337 & 3 \\
2 & -45.183831512387 & 0.122766222025 & 5 \\
3 & -44.873713217929 & 0.125865689868 & 18 \\
\hline\hline
\end{tabular}
\end{minipage}
\hfill
\begin{minipage}{0.47\textwidth}
\centering
\setlength{\tabcolsep}{4pt}
\begin{tabular}{l l l l}
\hline\hline
\multicolumn{4}{c}{$U/t=12$} \\
\hline
\multicolumn{1}{c}{$m$} & \multicolumn{1}{c}{$E_m/t$} & \multicolumn{1}{c}{$D_m$} & \multicolumn{1}{c}{$g_m$} \\
\hline
0 & -101.992223396705 & 0.027786853278 & 1 \\
1 & -101.843531263719 & 0.027289043346 & 3 \\
2 & -101.542398546613 & 0.026232122498 & 5 \\
3 & -101.303689197024 & 0.024849760390 & 12 \\
\hline\hline
\end{tabular}
\end{minipage}
\end{table}

Since $g_m$ and $e^{-\beta(E_m - E_0)}$ are positive numbers, Eq.~(\ref{eq:TrucForD}) clearly reveals that, provided $D_m>D_0$ holds for all involved excited states ($1\le m\le M$), then one obtains $D(T)>D_0$ and correspondingly $(\partial D/\partial T)_U>0$ near $T=0$ (and conversely, $D_m<D_0$ leads to $D(T)<D_0$ and hence $(\partial D/\partial T)_U<0$). This conclusion is more evident from a further approximation of Eq.~(\ref{eq:TrucForD}), providing $\sum_{m^{\prime}=1}^M g_{m^{\prime}} e^{-\beta(E_{m^{\prime}} - E_0)}\ll 1$, as
\begin{equation}\begin{aligned}
\label{eq:TrucForD1}
\langle D\rangle_{\rm M}^{(1)}(T) \approx D_0 + \sum_{m=1}^M g_m (D_m - D_0) e^{-\beta(E_m - E_0)}.
\end{aligned}\end{equation}
This formula is valid at low temperature limit, and it obviously shows $(\partial D/\partial T)_U>0$ if $D_m>D_0$ [and $(\partial D/\partial T)_U<0$ if $D_m<D_0$]. 

\begin{figure}[t]
\centering
\includegraphics[width=0.66\textwidth]{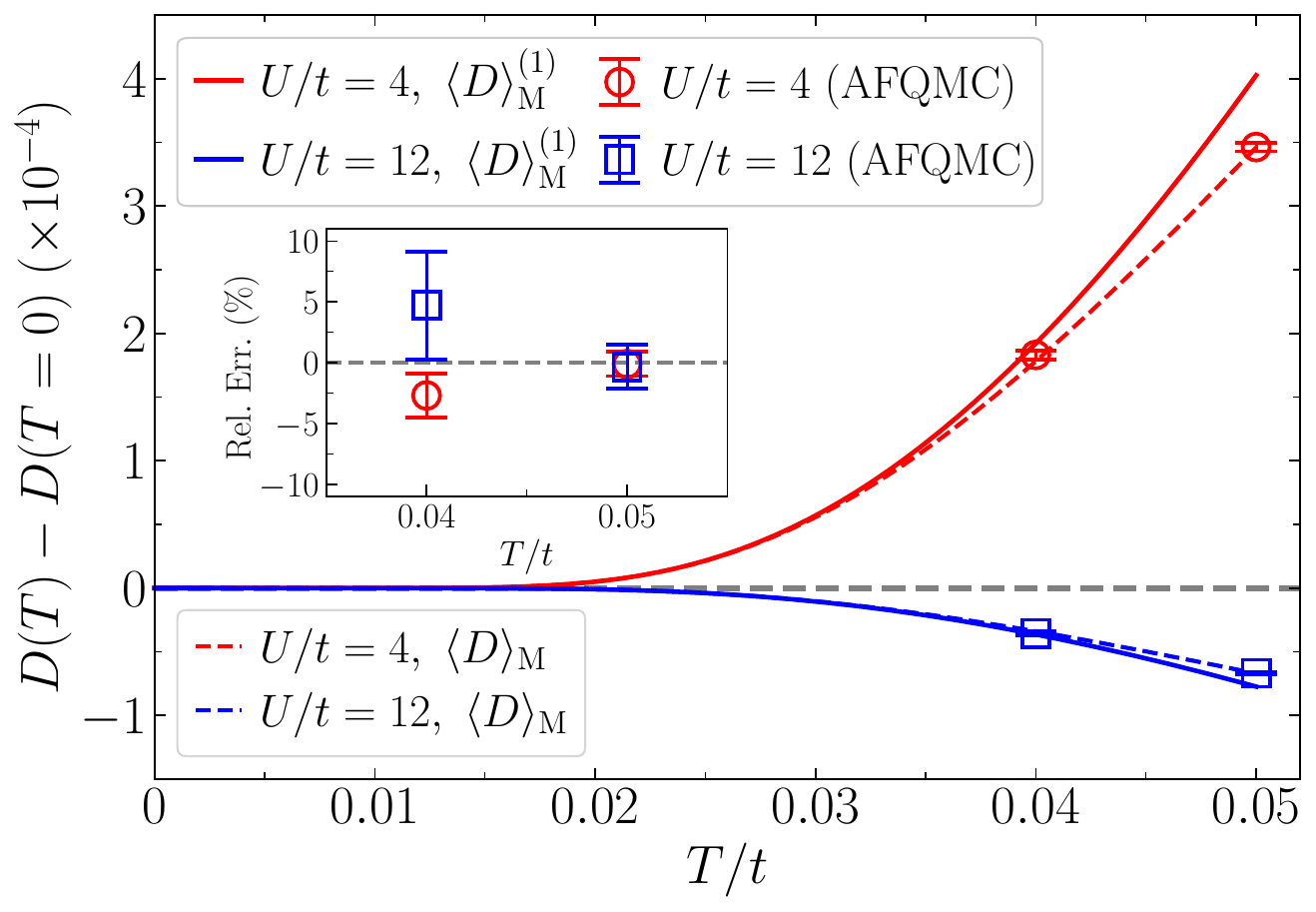}
\caption{
Comparison results of $[D(T)-D(T=0)]$ as a function of $T/t$ in the range $0\le T/t\le 0.05$ for $U/t=4$ and $12$ from both ED (solid and dashed lines) and AFQMC (symbols) calculations. The ED results obtained from both Eq.~\eqref{eq:TrucForD} (solid lines) and Eq.~\eqref{eq:TrucForD1} (dashed lines) are shown. The inset shows the relative error between ED via Eq.~\eqref{eq:TrucForD} and AFQMC results.
}
\label{fig:FigS2DT}
\end{figure}

\begin{figure}[t]
\centering
\includegraphics[width=0.66\textwidth]{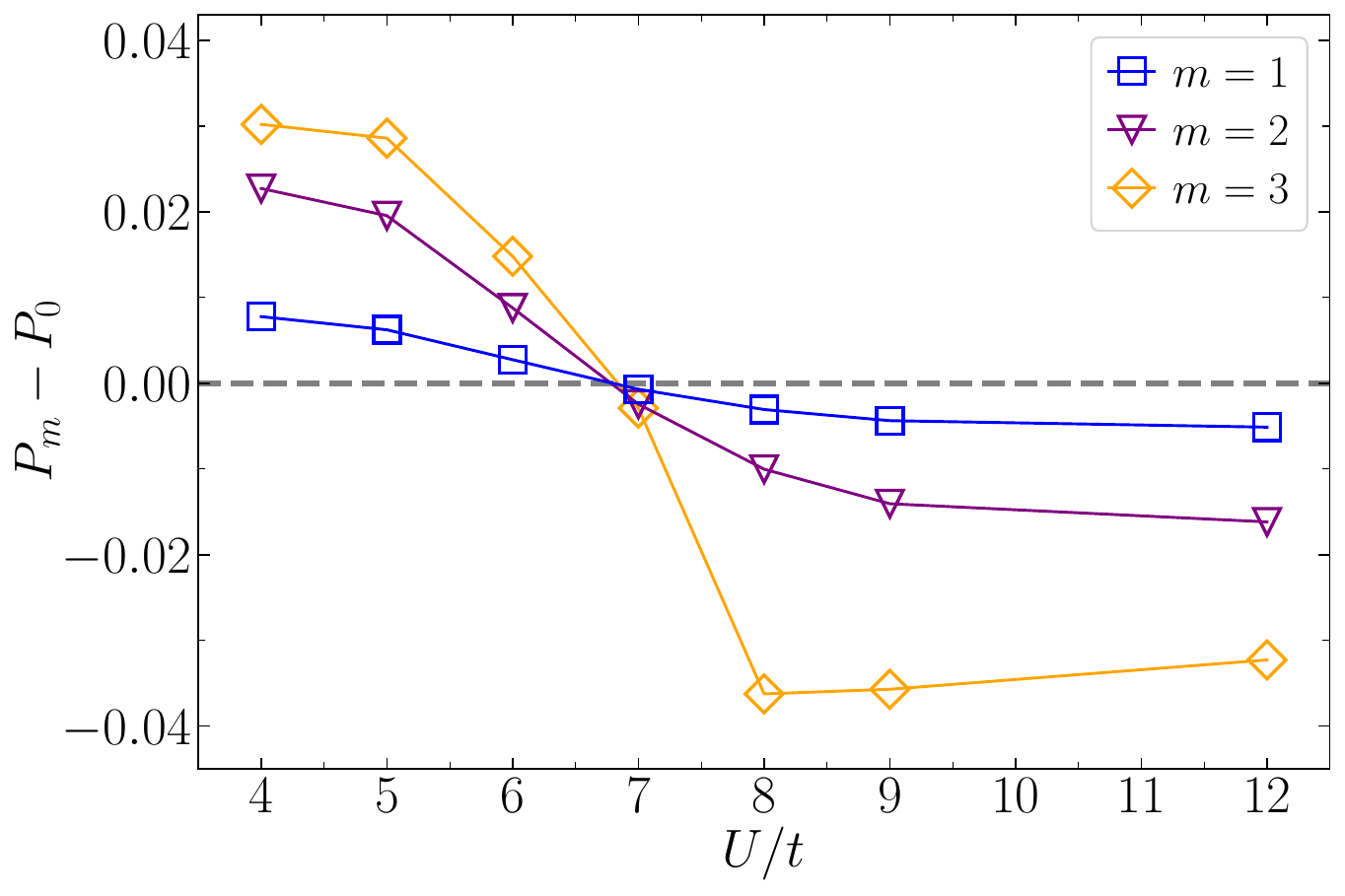}
\caption{
ED results of $(P_m-P_0)$ as a function of $U/t$, with $P_m$ being the accumulated weight of Fock states containing at least one doubly occupied site in the $m$-th excited state wavefunction, and $P_0$ the ground-state counterpart.
}
\label{fig:FigS3Pm}
\end{figure}

In Fig.~2 inset in the main text, we compute the $D(T)$ results via Eq.~(\ref{eq:TrucForD}) for $U/t=4$ and $U/t=12$, in the temperature range of $0\le T/t\le 0.05$. The calculation uses the ED results of $(E_m,D_m,g_m)$ from $4\times4$ system under PBC, where $E_m$ is the eigenenergy, $D_m$ the double occupancy, and $g_m$ the degeneracy, of the $m$-th excited state. These $(E_m,D_m,g_m)$ results are listed in Table~\ref{Tab02EDLevels}. For the half-filled Hubbard model, the ground state lies in the $(N_{\uparrow},N_{\downarrow})=(8,8)$ subspace [$N_{\uparrow}$ ($N_{\downarrow}$) denotes the number of spin-up (spin-down) electrons], and the excited states needed in our calculations involve the $(N_{\uparrow},N_{\downarrow})=(7,9)$ and $(9,7)$ subspaces. For the temperature range $0\le T/t\le 0.05$, we find that applying $M=3$ in Eq.~(\ref{eq:TrucForD}) can already render the approximation induced by the truncation negligible for all the interaction strengths involved ($4\le U/t\le 12$). To assess the difference between Eqs.~(\ref{eq:TrucForD}) and~(\ref{eq:TrucForD1}), we compute $D(T)$ via both formulas and compare with AFQMC results, as illustrated in Fig.~\ref{fig:FigS2DT}. The $D(T)$ results from both Eqs.~(\ref{eq:TrucForD}) and~(\ref{eq:TrucForD1}) confirm the sign conditions, i.e., $(\partial D/\partial T)_U>0$ for $U/t=4$ and $(\partial D/\partial T)_U<0$ for $U/t=12$, for the whole plotted range of $T/t$. Moreover, the results from Eqs.~(\ref{eq:TrucForD}) and~(\ref{eq:TrucForD1}) show observable differences for both values of $U/t$ at $T/t\ge 0.04$, and the those from Eq.~(\ref{eq:TrucForD}) agree well AFQMC, as indicated by the relative errors in Fig.~\ref{fig:FigS2DT} inset, which in turn validates the truncation with $M=3$ in Eq.~(\ref{eq:TrucForD}).

In the main plot of Fig.~2 in the main text, we present the ED results of $(D_m-D_0)$ and reveal the underlying Slater versus Mott physics. Here, in Fig.~\ref{fig:FigS3Pm}, we show the corresponding results of $(P_m-P_0)$ as a function of $U/t$ (for $m=1,2,3$), with $P_m$ being the accumulated weight of Fock states containing at least one doubly occupied site in the $m$-th excited state wavefunction, and $P_0$ the ground-state correspondence. We find that $(P_m-P_0)$ is positive for $U/t\le 6$ and negative for $U/t\ge 7$, closely resembling the behavior of $(D_m-D_0)$. This further confirms the Slater-versus-Mott physics revealed by $(D_m-D_0)$.

\section{IV. Finite-size effects of \texorpdfstring{$D(T)$}{} around the crossover}
\label{sec:FiniteSize}

In Fig.~3 of the main text, we have presented AFQMC results of $[D(T)-D(T=0)]$ around the Slater-Mott crossover boundary $U_{\rm cross}/t=4.0(2)$ from TABC calculations with $L=20$. The finite-size effects are verifed to be negligible in those data. In this section, we present additional AFQMC results in smaller systems to explicitly illustrate the finite-size effects in $D(T)$.

\begin{figure}[H]
\centering
\includegraphics[width=0.99\textwidth]{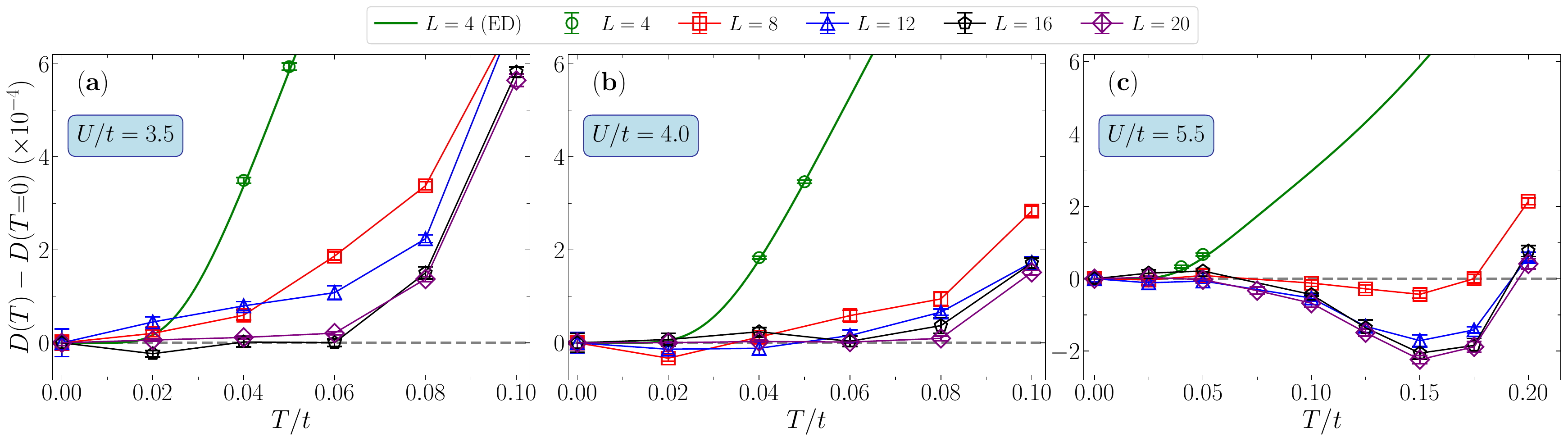}
\caption{
Numerical results of $[D(T)-D(T=0)]$ as a function of $T/t$, for (a) $U/t=3.5$, (b) $U/t=4.0$, and (c) $U/t=5.5$, with $L=4$, $8$, $12$, $16$, and $20$. For $L=4$, the ED results are also included as a benchmark. The AFQMC results with $L=4\sim16$ are under periodic boundary conditions, while the $L=20$ data are simply reproduced from Fig.~3 in the main text. The $T=0$ results (denoted as $D(T=0)$) are obtained from ground-state AFQMC simulation.}
\label{fig:FigS4CompareL}
\end{figure}

In Fig.~\ref{fig:FigS4CompareL}, we compare the numerical results of $[D(T)-D(T=0)]$ as a function of $T/t$ for $U/t=3.5$, $4.0$, and $5.5$, with $L=4\sim 20$. For $L=4$, we have also included the ED results, computed using Eq.~(3) in the main text (similar to the inset of Fig.~2), as benchmarks. The AFQMC results from $L=4\sim16$ are obtained under periodic boundary conditions (PBC), while the $L=20$ data are reproduced from Fig.~3 in the main text. For all three interactions, the $L=4$ results from ED confirm $(\partial D/\partial T)_U>0$ near $T=0$, manifesting the Slater behavior in the small system. At $U/t=3.5$ [see Fig.~\ref{fig:FigS4CompareL}(a)], the $D(T)$ results gradually converge as $L$ increases, with the positive derivative $(\partial D/\partial T)_U>0$ unchanged. Note that the simulation results under PBC with $L=4\sim16$ show some fluctuations at low temperatures, while the TABC results from $L=20$ is generally more reliable. At $U/t=4.0$ [see Fig.~\ref{fig:FigS4CompareL}(b)], the $D(T)$ results show convergence at $0<T/t<0.06$ for $L\ge12$, which evidently show that $D(T)$ remains almost constant with statistical uncertainties in this low-$T$ regime, indicating $(\partial D/\partial T)_U\simeq0$. This is consistent with the crossover boundary. At $U/t=5.5$ [see Fig.~\ref{fig:FigS4CompareL}(c)], in contrast, an evident sign change in $(\partial D/\partial T)_U$ can be observed across $L=4$ to $L=8$, manifesting the finite-size effects. Moreover, $D(T)$ results almost converge for $L\ge12$ in the whole plotted range of $T/t$, yielding $(\partial D/\partial T)_U<0$ near $T=0$. 

As discussed in the main text, the ED results from $L=4$ shown in Fig.~2 indicate the Slater-Mott crossover boundary located at $U/t=6\sim7$. According to the numerical results in Fig.~\ref{fig:FigS4CompareL}(c), $U/t=5.5$ belongs to the Slater regime at $L=4$ as $(\partial D/\partial T)_U>0$, whereas it moves into the Mott regime at $L\ge8$ as $(\partial D/\partial T)_U<0$. This shows that the crossover boundary exhibits finite-size effect, from $U/t=6\sim7$ at $L=4$ to the AFQMC result $U_{\rm cross}/t=4.0(2)$ in the thermodynamic limit.

\section{V. Temperature dependence of the kinetic energy}
\label{sec:KineticEnergy}

Previously, the temperature dependence of the kinetic energy was used to characterize the Slater-Mott crossover, applying dynamical mean-field theory~\cite{Pruschke2003,Borejsza2003,Fratino2017,Wang2019} and diagrammatic Monte Carlo (DiagMC) method~\cite{Kim2020}. Specifically, in Ref.~\cite{Kim2020}, $(\partial e_K/\partial T)_U<0$ at low temperatures was proposed as a signature of the Slater insulator. In this section, we examine the temperature dependence of the kinetic energy density, computed as $e_K=N_s^{-1}\sum_{\mathbf{k}\sigma}\varepsilon_{\mathbf{k}}\langle\hat{c}_{\mathbf{k}\sigma}^{\dagger}\hat{c}_{\mathbf{k}\sigma}^{}\rangle$, across the Slater-Mott crossover.

\begin{figure}[H]
\centering
\includegraphics[width=0.66\textwidth]{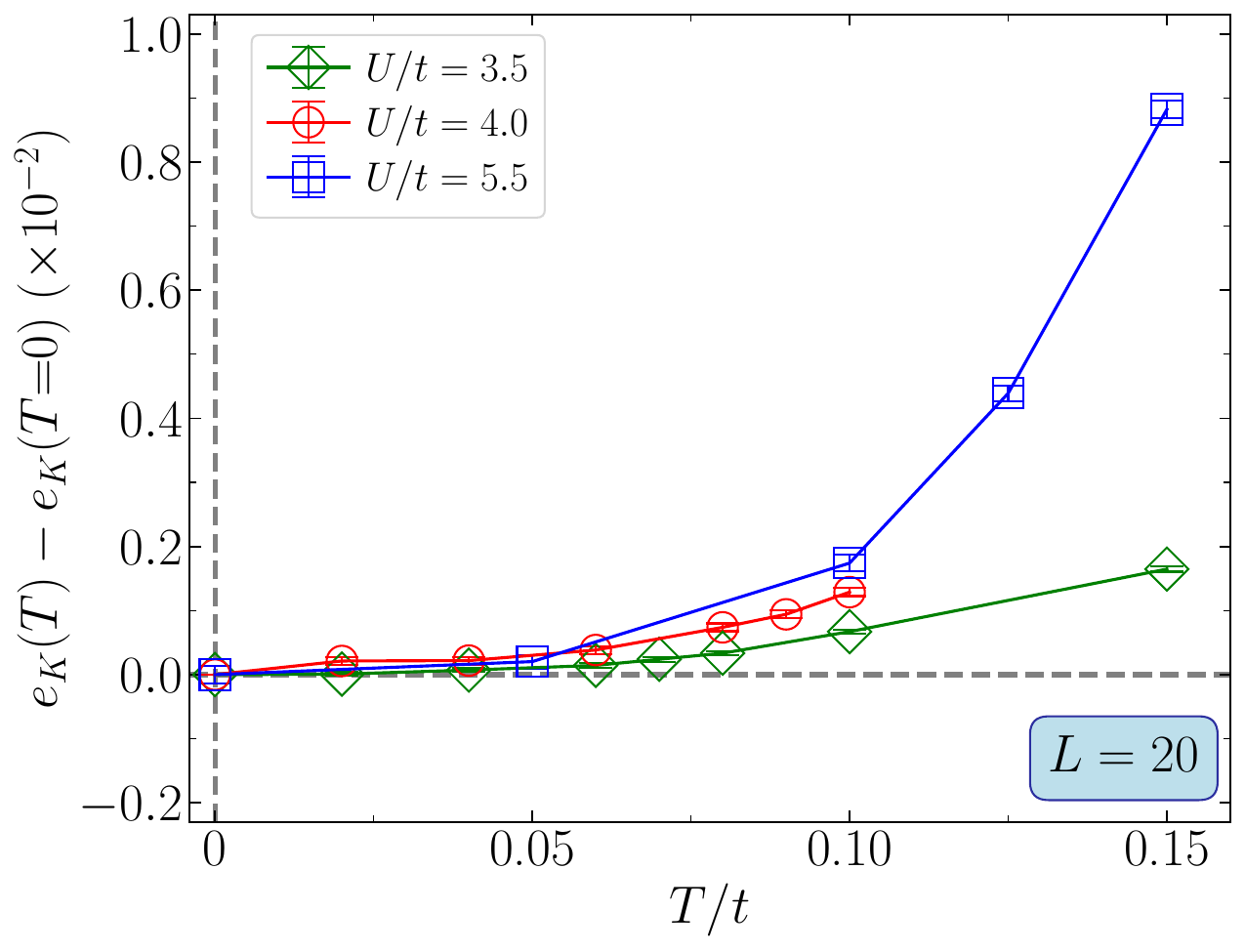}
\caption{
AFQMC results of the kinetic energy density difference, denoted $e_K(T)-e_K(T{=}0)$, as a function of $T/t$ at $U/t=3.5$, $4.0$, and $5.5$ with $L=20$ under TABC. The $T=0$ and $T>0$ results are accordingly computed using ground-state and finite-temperature AFQMC methods, and exhibit negligible finite-size effects.
}
\label{fig:FigS5KwithT}
\end{figure}

In Fig.~\ref{fig:FigS5KwithT}, we present our AFQMC results of $[e_K(T)-e_K(T{=}0)]$ at $U/t=3.5$, $4.0$, and $5.5$ with $L=20$ under TABC. The kinetic energy density increases monotonically with temperature for all three interaction strengths, indicating $(\partial e_K/\partial T)_U>0$ throughout. In particular, at $U/t=3.5$, the results in Ref.~\cite{Kim2020} would suggest $(\partial e_K/\partial T)_U<0$ below some temperature between $T/t=0.125$ and $0.225$, yet our results show $(\partial e_K/\partial T)_U>0$ over the entire temperature range down to $T=0$. Since $(\partial e_K/\partial T)_U>0$ holds across both the Slater ($U/t=3.5$) and Mott ($U/t=5.5$) regimes, the kinetic energy density does not provide a clear criterion for distinguishing the Slater and Mott regimes. We attribute this discrepancy to the relatively large statistical uncertainties in the low-temperature data of Ref.~\cite{Kim2020} (and actually the $(\partial e_K/\partial T)_U<0$ behavior was never explicitly exposed in Ref.~\cite{Kim2020}). Our AFQMC results, obtained with substantially improved statistics, do not support $(\partial e_K/\partial T)_U<0$ in the Slater regime. Furthermore, from a physical respective, given that the half-filled Hubbard model has an insulating ground state for any $U>0$, the kinetic energy $e_K$ is expected to increase upon heating in the low-$T$ regime, as thermal fluctuations enhance the probability of hopping processes. This leads to $(\partial e_K/\partial T)_U>0$. Our numerical results in Fig.~\ref{fig:FigS5KwithT} are consistent with this physical picture.

\section{VI. The Slater-Mott crossover in the half-filled 3D Hubbard model}
\label{sec:CorrLength3D}

In this section, we further present our calculation results for the Slater-Mott crossover in the half-filled 3D Hubbard model. We have computed both the double occupancy $D$ and AFM correlation length $\xi$. However, we find that, $D$ exhibits only tiny variations in the low-temperature regime (such as $T/t\le 0.15$) at $U\sim U_{\rm cross}^{3D}$, which is likely to arise from being deep inside the AFM long-range ordered phase. Consequently, it is very difficult to resolve the sign of $(\partial D/\partial T)_U$ near $T=0$ directly from the $D(T)$ data. Therefore, we mainly focus on the AFQMC results of $\xi$ to determine the crossover boundary. 

\begin{figure}[H]
\centering
\includegraphics[width=0.99\textwidth]{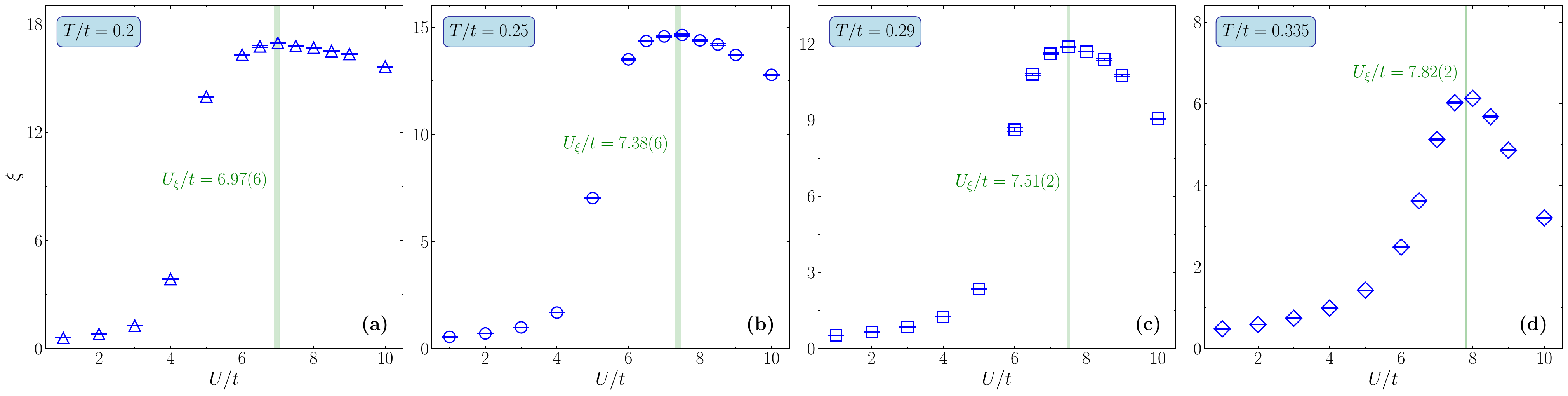}
\caption{
AFM correlation length $\xi$ as a function of $U/t$ for the half-filled 3D Hubbard model on the simple cubic lattice at (a) $T/t=0.2$, (b) $T/t=0.25$, (c) $T/t=0.29$, and (d) $T/t=0.335$, using $L=12$ AFQMC data under PBC. The green shaded band in each panel marks the peak position $U_\xi/t$.
}
\label{fig:FigS6CorrLength3D}
\end{figure}

\begin{figure}[H]
\centering
\includegraphics[width=0.66\textwidth]{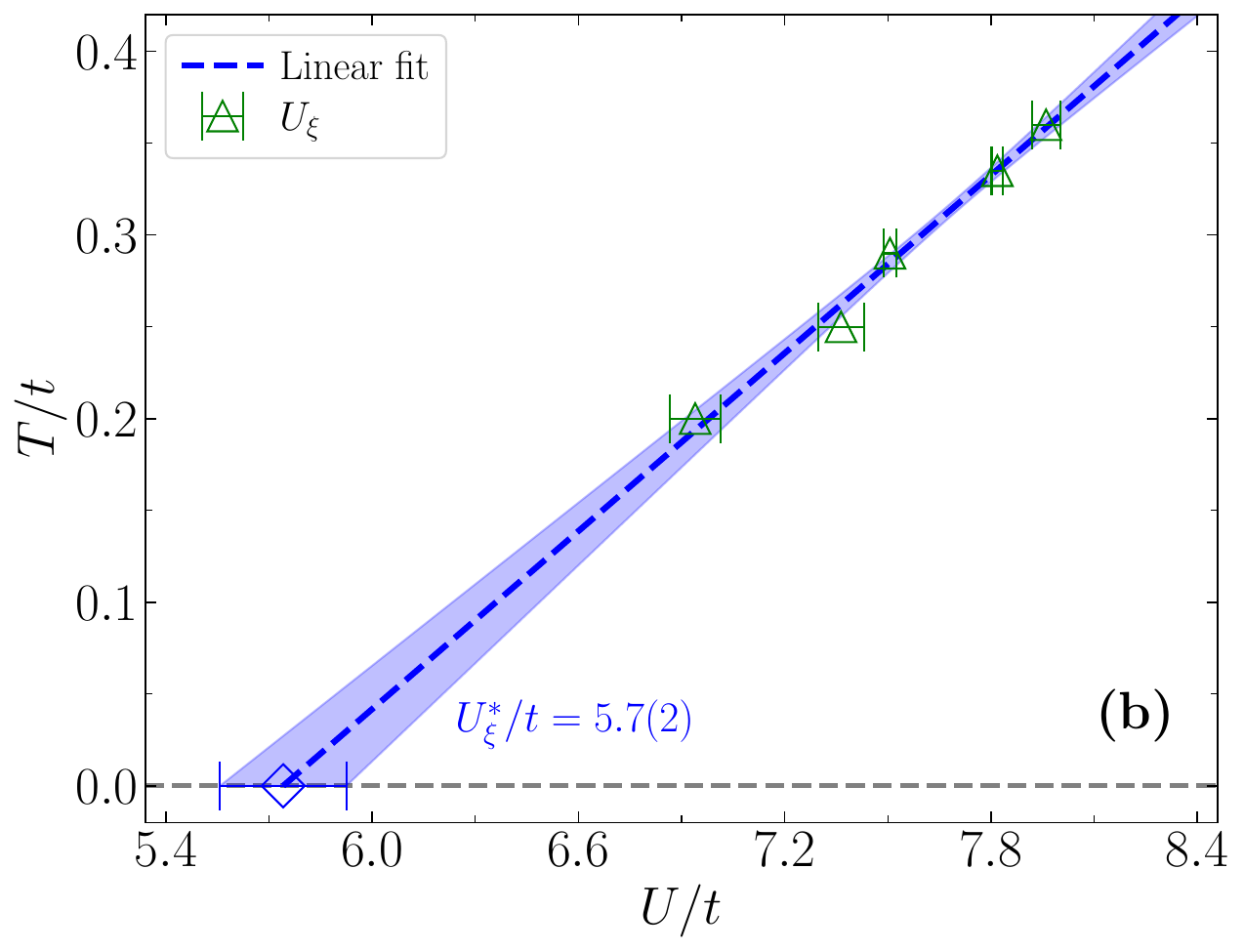}
\caption{
Linear extrapolation of the AFM correlation length peak position $U_\xi/t$ to $T=0$ for the half-filled 3D Hubbard model on the simple cubic lattice, using $L=12$ AFQMC data under PBC. The dashed line shows the linear fit, with the shaded region indicating the fitting uncertainty.
}
\label{fig:FigS7CorrLengthExtrap}
\end{figure}

In Fig.~\ref{fig:FigS6CorrLength3D}, we show the AFQMC data of $\xi$ as a function of $U/t$ at four temperatures from $L=12$ simulations under PBC. Although $\xi$ diverges in the ordered phase in the thermodynamic limit, finite-size simulations still yield large but finite values, as seen near the peak positions in Figs.~\ref{fig:FigS6CorrLength3D}(a)--(c). The green shaded band in each panel marks the peak position $U_\xi$, i.e., the interaction strength at which $\xi$ reaches the maximum. Regardless of the significant finite-size effects in $\xi$ itself, we have verified that $U_{\xi}$ instead converges quickly with $L$, and the $U_{\xi}$ values obtained from $L=12$ in Fig.~\ref{fig:FigS6CorrLength3D} already converge within the statistical uncertainty. 

In Fig.~\ref{fig:FigS7CorrLengthExtrap}, we show the linear extrapolation to $T=0$ for the $U_\xi$ values obtained in Fig.~\ref{fig:FigS6CorrLength3D}, which yields $U_\xi^*/t=5.7(2)$ as the crossover boundary. This illustrates that the approach developed for the 2D model in the main text can be readily extended to other systems hosting the Slater-Mott crossover.

\end{document}